\documentclass[twocolumn,secnumarabic,amssymb, nobibnotes, aps, prd]{revtex4-2}

\newcommand{\env}[1]{\texttt{#1}}
\usepackage{graphicx}
\usepackage{tabularx}
\usepackage{subcaption} 
\usepackage{array} 
\usepackage{color,soul}
\usepackage{ar}
\usepackage{xspace}

\begin{document}

\title{Effects of Aspect Ratio on Rolling and Twisting Foils}%

\author{Andhini N. \surname{Zurman-Nasution}}%
\author{Bharathram Ganapathisubramani}
\author{Gabriel D. Weymouth}
\email[ ]{A.N.Zurman-Nasution@soton.ac.uk}
\affiliation{University of Southampton, Boldrewood Innovation Campus, Burgess Road, Southampton, SO16 7QF.}
\date{\today}%

\begin{abstract}
Flapping flight and swimming are increasingly studied both due to their intrinsic scientific richness and their applicability to novel robotic systems. Strip theory is often applied to flapping wings, but such modeling is only rigorously applicable in the limit of infinite aspect ratio (\AR) where the geometry and kinematics are effectively uniform. This work compares the flow features and forces of strip theory and three dimensional flapping foils, maintaining similitude in the rolling and twisting kinematics while varying the foil \AR. We find the key influence of finite \AR\xspace and spanwise varying kinematics is the generation of a time-periodic spanwise flow which stabilizes the vortex structures and enhances the dynamics at the foil root. An aspect-ratio correction for flapping foils is developed analogous to Prandtl finite wing theory, enabling future use of strip theory in analysis and design of finite aspect ratio flapping foils.
\end{abstract}

\maketitle
\tableofcontents

\section{Introduction}

Biologically inspired robotic systems offer benefits in their flexibility, maneuverability, energy savings, and speed. However, development of bio-inspired micro air vehicles (MAVs) and autonomous underwater vehicles (AUVs) requires a detailed understanding of unsteady three-dimensional fluid mechanics of propulsive flapping at intermediate Reynolds-number ($Re\approx1000-10000$). Because of the cost and complexity of three-dimensional (3D) simulations, most propulsive flapping studies are conducted using a strip theory approach which assumes a simplified two-dimensional (2D) flow across the wing sections. However, this approach can lead to poor engineering predictions for the forces and efficiency due to the inherently 3D nature of the geometry, motion, and flow structures \cite{Mittal1995}. 

The key 3D fluid dynamic effects on a finite flapping foil undergoing 3D motion are spanwise flow, tip vortices, and 3D instabilities of the vortex wake. Spanwise flow promotes the fluid dynamic interaction between sections of the wing which can result in high lift generation \cite{Hong2008,Ellington1996, Maxworthy2007}. The exact mechanism of this high lift on bird or insect wings is still highly debated between the spanwise flow \cite{Ellington1996,Wong2017,Eldredge2019} and downwash-induced flow caused by the tip \cite{Lentink2009,Birch2001,Lu2008}. In addition, the tip of a finite foil definitely produces non-planar wakes which have been shown to influence stall on an impulsively translated flat plates \cite{Taira2009}, pitching flat plates \cite{Hartloper2013}, revolving wings \cite{Harbig2013,Jardin2017,Kim2010,Garmann2014} and a finite foil undergoing 2D kinematics \cite{Dong2005}. Finally, even on an infinite wing with no tip effect or mean spanwise flow, 3D evolution of the vortex wake limits strip theory applications. \citet{zurman2020} showed there is only exist a narrow band of kinematics, for instance at Strouhal number $\approx 0.15-0.45$ for heaving-pitching motion, where the wake remains two-dimensional and 2D force predictions are accurate. In this 2D range, the motion of the foil stabilize the spanwise-perturbed 3D structures found in the separated wake of a stationary foil, but 3D structures reappear when the amplitude of motion increases the shed vortex circulation above a 2D viscous stability limit. This secondary instability along spanwise-aligned vortex tubes also causes the transition from 2D to 3D flow in circular cylinder wakes \cite{Karniadakis1992,Williamson2006}. Despite the importance of these effects, wake analysis for finite foils undergoing 3D kinematics is still rare because of the greatly reduced simulation cost when using 2D strip theory.

The key geometric parameter in the flow and forces on 3D flapping foil is the aspect ratio, \AR, defined as the square of total wing span length over the planform area. \citet{Chopra1974} provided potential flow calculations indicating that decreasing \AR\xspace would reduce thrust coefficient and efficiency. On the contrary, move recent experiments by \citet{Fu2015} show that the circulation over the tip velocity increases with \AR\xspace but drops at $\textrm{\AR}=4$, potentially explaining why many insects have low \AR\xspace wings. These contradicting results show the need for simulations detailing the \AR\xspace influence on the flow and forces of wings undergoing 3D kinematics. For a finite wing undergoing simple 2D pitching motion, pressure and thrust coefficient data are found to scale well with $1/(1+c' \textrm{\AR})$ where $c'$ is a constant \cite{Green2008,Ayancik2019}. This is similar to Prandtl finite wing theory although \citet{Smits2019} emphasizes that this added mass scaling is somehow misleading because it neglects the circulatory forces. In addition, these studies and others such as \cite{Calderon2013,Yilmaz2011} which use purely 2D kinematics cannot study the contributions of 3D kinematics on the flow features and force scaling.

The ability to make 3D predictions from 2D simulations using analogous concepts to Prandtl's finite wing theory is an intriguing approach for speeding up computations, but it must be based on a strong understanding of the influence of \AR\xspace on the unsteady 3D flow. In this work, we provide a detailed analysis using 3D simulations of rolling and twisting foils; a combination of 3D kinematics that imitates the natural flapping motion of animal's flippers \cite{Clark1978,Davenport1984} but can also be directly mapped to simple pitch and heave motion of each section. The addition of 3D twist to the foil kinematics also relates the current work to previous studies which indicate improved lift \cite{Vanburen2018_1,Cleaver2014} and thrust coefficients \cite{Cleaver2016,Barnes2013} for flexible foils. The instantaneous and phase-averaged vortex structures in the tip flow and wake, the spanwise flow along the foil, and the unsteady sectional forces on the foil are studied while aspect ratio is varied. Finally, we relate the sectional forces on the 3D foils to Prandtl finite-wing theory in order to scale the 2D strip theory force predictions onto 3D flapping wings.

\section{Methodology}
\subsection{Geometry and kinematics}

\begin{figure*}[ht!]
\centering
	\begin{subfigure}[b]{1\textwidth}
		\includegraphics[width=.6\textwidth]{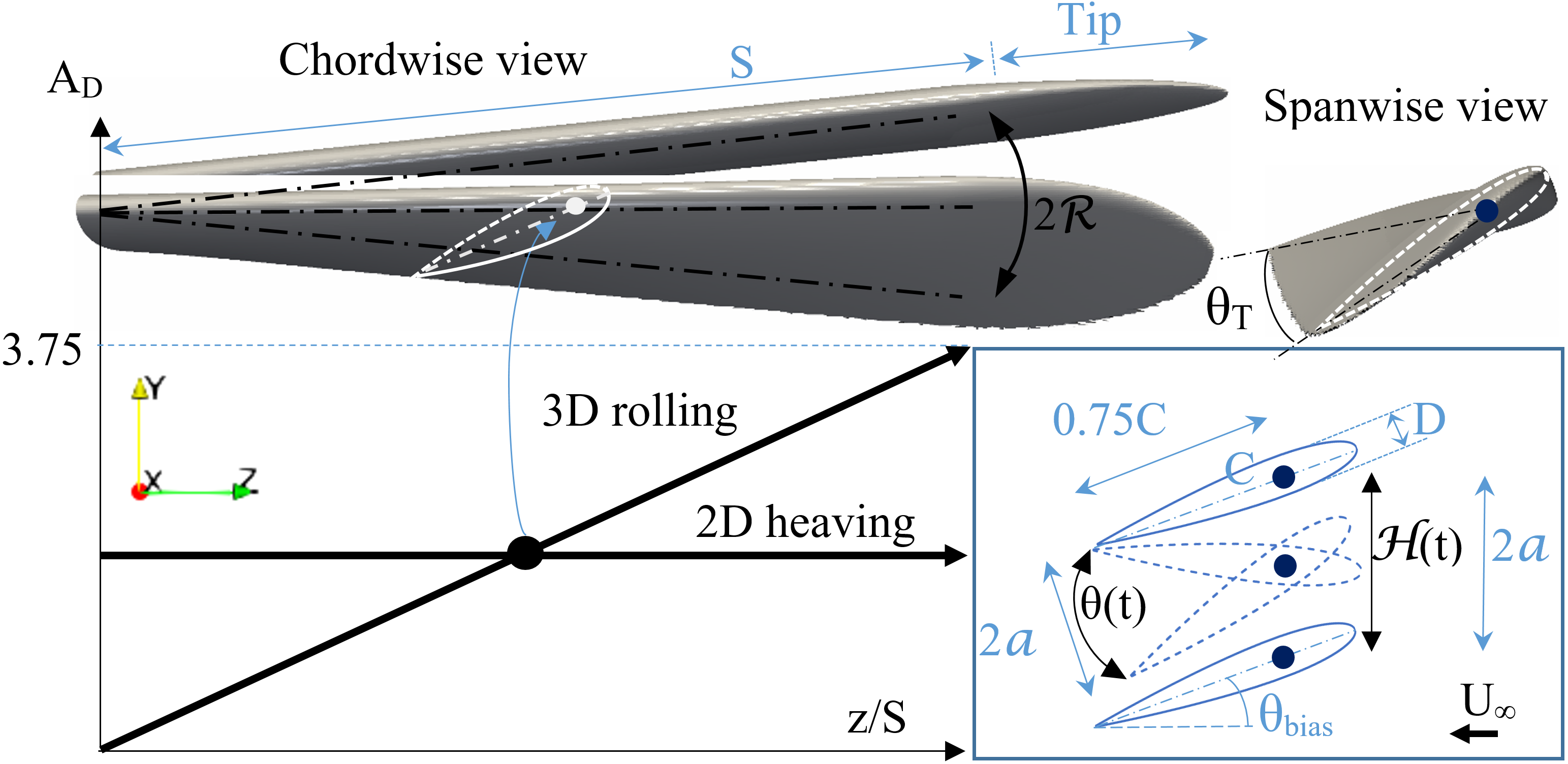}
		\subcaption{}
	\end{subfigure}
    \begin{subfigure}[b]{1\textwidth}
		\includegraphics[width=0.85\textwidth]{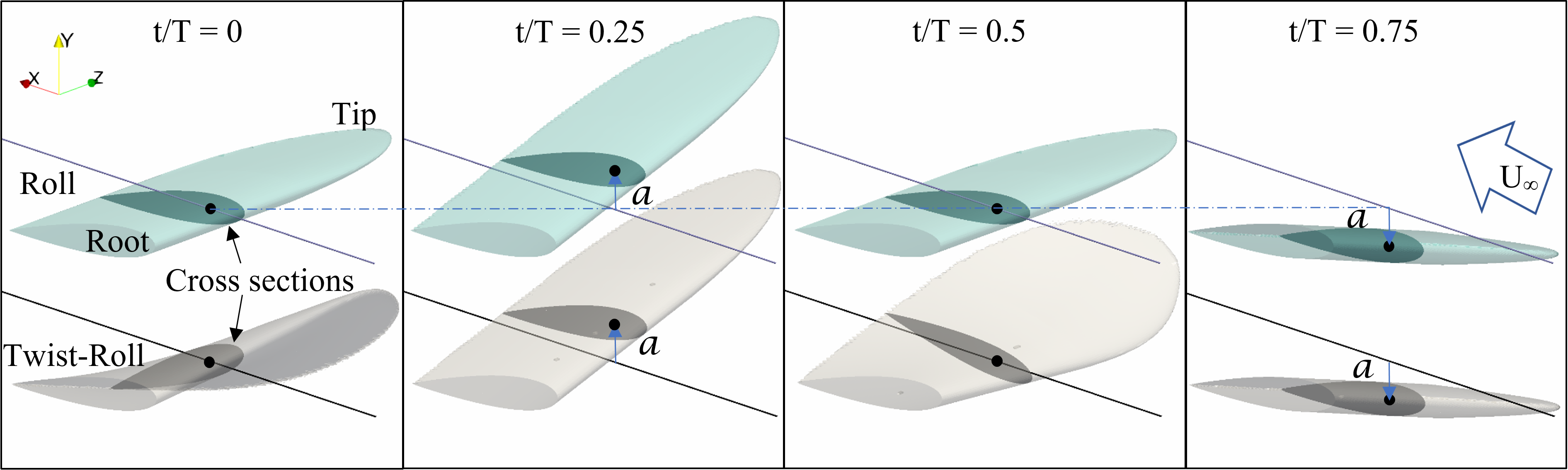} 
		\subcaption{}
	\end{subfigure}
\caption{(a) Kinematic illustration for a 3D foil undergoing rolling and twisting motions whose cross sections are comparable to 2D foil experiencing linearly increased heave and pitch amplitude along the span. (b) Illustration of 4 phases in one cycle period (T) for pure rolling kinematic and twisting-rolling combination.}
\label{fig:kine}
\end{figure*}

This study uses a NACA0016 foil with chord length $C$, thickness \(D=0.16C\), pivot point at $0.25C$ from the leading edge, a rectangular planform section with span length $S$, and a tapered elliptic tip with a length of $1C$ as shown in Fig. \ref{fig:kine}. We will characterize the aspect ratio by $S/C$ which is varied from \(1\ldots6\). The foil is placed in a uniform flow with constant inflow speed $U_\infty$ and a fluid density $\rho$ and the chord-based Reynolds number $Re=U_\infty C/\nu=5300$.

The 3D foil is prescribed to move in either pure roll or a combination of roll and twist, where roll is defined as rigid rotation around the streamwise axis \(x\) and twist is rotation around the spanwise axis \(z\) with linearly increasing amplitude towards the tip, Fig. \ref{fig:kine}. 

In a strip theory approach, these 3D kinematics are converted into 2D kinematics for each section along the span. The 2D sectional kinematics can be linearized for small amplitude roll to simple heave $\mathcal{H}(t)$ and pitch $\theta(t)$ motion in each spanwise plane, with functional form
\begin{equation} \label{heaving}
	\mathcal{H}(t) = a \sin(2 \pi f t)
\end{equation}
\begin{equation} \label{pitching}
	\theta(t) = \theta_{0} \sin(2 \pi f t + \psi) + \theta_{\text{bias}}
\end{equation}
\begin{equation}\label{pitching_amp}
	\theta_{0} = \sin^{-1}\big(a/(0.75C)\big)
\end{equation}
where  \(f\) is the cycle frequency of flapping and \(a\) and \(\theta_{0}\) are the amplitudes of heave and pitch at each section. The term $\theta_{\text{bias}}=10^\circ$ is a pitch bias angle used to ensure non-zero average mean lift as in \citet{zurman2020} and the phase difference is set to $\psi=90^\circ$ to maximize performance \cite{Isogai1999}. In order to isolate the amplitude (which varies section by section) from the frequency, these motions are characterized as
\begin{equation} \label{StD_AD}
	St_{D}=\frac{Df}{U_\infty} \qquad A_{D}=\frac{2a}{D}
\end{equation}
where \(St_{D}\) is the thickness-based Strouhal number, \(A_{D}\) is the thickness-based amplitude and in this work we fix \(St_{D}=0.3\). Combining these two gives the commonly used amplitude-based Strouhal number \(St_A= St_D \cdot A_D\) \cite{Triantafyllou1991} which is proportional to the maximum heave velocity on a section $\dot{\mathcal{H}}_{max}=2\pi fa$ scaled by the incoming flow velocity. 

Strip theory assumes that both the geometry and the motions change slowly along the span. The geometric constraint is unavoidably violated at the tip, but the kinematic condition is still a concern over the rectangular region of the planform. Slowly varying kinematics are contingent on large aspect ratio; a relationship that can be made explicit by noting that the pitch amplitude at each section is \(\theta_{0}=\theta_T z/S\) where \(\theta_T\) is the twist amplitude at the foil tip, Fig. \ref{fig:kine}. Therefore the scaled spanwise derivative of the 2D strip theory kinematics is \(C\partial_z\theta_{0}= \theta_T C/S\). The larger the aspect ratio, the smaller this derivative, and the better strip-theory should apply. Similarly, for small roll amplitude arc-length \(\mathcal{R}\), the sectional heave amplitude is \(a=\mathcal{R} z/S\) and therefore \(\partial_z a=\mathcal{R}/S\). To focus on the aspect ratio effect, we prescribe a constant roll amplitude \(\mathcal{R} = 0.3C\) from which all other amplitudes are determined. 

\begin{figure}[hbt!]
	\centering
	\includegraphics[trim=0 0 0 1.8cm, clip=true, width=0.4\textwidth]{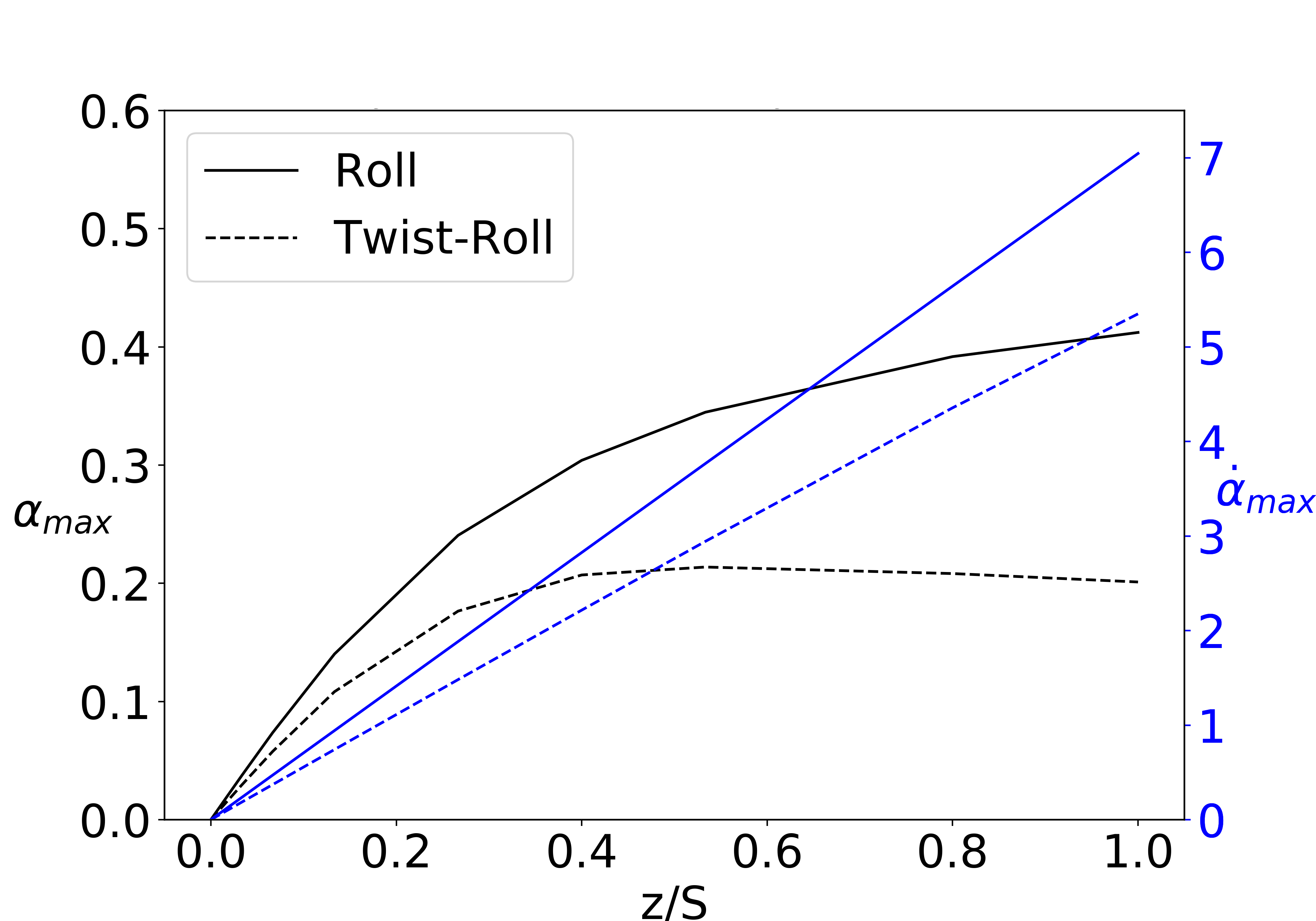}
\caption{The gradient of maximum angle of attack  $\alpha_{max}$ ($\pi$ rad) towards the tip is positive with linear rate $\dot{\alpha}_{max}$ ($\pi$ rad/t).}
\label{fig:aoa}
\end{figure}  

Another way to measure the spanwise change in 2D kinematics along the foil, used recently by \cite{Wong2017}, is the sectional angle-of-attack $\alpha$
\begin{equation}\label{aoa_eq}
	\alpha(t) = \tan^{-1}\big(\dot{\mathcal{H}(t)}/U_{\infty})\big)-\theta(t)
\end{equation}
Because $\dot{\mathcal{H}(t)}$ increases towards the tip, so do the sectional angle-of-attack $\alpha$ and it's time derivative $\dot{\alpha}$. Fig. \ref{fig:aoa} shows $\dot{\alpha}_{max}$ increases linearly along the span, while trigonometry causes the change in $\alpha_{max}$ to taper off near the tip. In either case, the spanwise derivative of these parameters depend inversely on $S/C$ and we show they strongly impact the flow three-dimensionality.

\subsection{Numerical methods}

The computational package used for this research is the Boundary Data Immersion Method (BDIM). \citet{Weymouth2011} introduced BDIM using a robust and efficient Cartesian grid with implicit LES (iLES) solver. BDIM uses analytic meta-equations for an immersed body in multi-phase flow with a smoothed interface domain using an integral kernel. BDIM has been validated, proved suitable for moving bodies and capable of resolving the flow at \(Re=10^5\) within 5\% error for thrust prediction \cite{Maertens2014}.

For computational consistency, the 2D and 3D simulations use the same solver. Symmetric conditions are enforced on both spanwise boundaries for the finite foil simulation. The domains extend from the pivot point to $4C$ at the front, $11C$ at the rear, $5.5C$ at the top and bottom. Meanwhile, the tip distance to maximum spanwise domain is $3.5C$. A grid convergence study with an infinite foil with span length of $3C$ and periodic spanwise boundary conditions and a 2D foil are used to verify the grid convergence for the lift and drag coefficients
\begin{equation} \label{Cl}
	 C_L=\frac{F_y}{0.5 \rho S_p U_{\infty}^2}	\qquad 	 C_D=\frac{F_x}{0.5 \rho S_p U_{\infty}^2}
\end{equation}
where \(F_x,F_y\) are the integrated fluid force in the global coordinate directions and \(S_p = CS+\frac\pi 4 C^2\) is the foil planform area. Fig \ref{fig:conv} in the appendix shows the standard deviation of the forces predicted using \(C/\Delta x=128\) are within 3\% of simulations with double the resolution in both 2D and 3D and this resolution is used for all results presented in this work.

\section{Results}\label{results}

In this section, we compare the flow-structure evolutions and forces of foil undergoing rolling and twisting-rolling motions over one cycle, across four aspect ratios. 

\subsection{Vortex structures}
\begin{figure}
	\begin{center}
	\begin{tabular}{|c|c|}
	\hline
	Roll	&  Twist-roll \\
	\hline
	\includegraphics[trim=0 0 0 0, clip=true, width=0.25\textwidth]{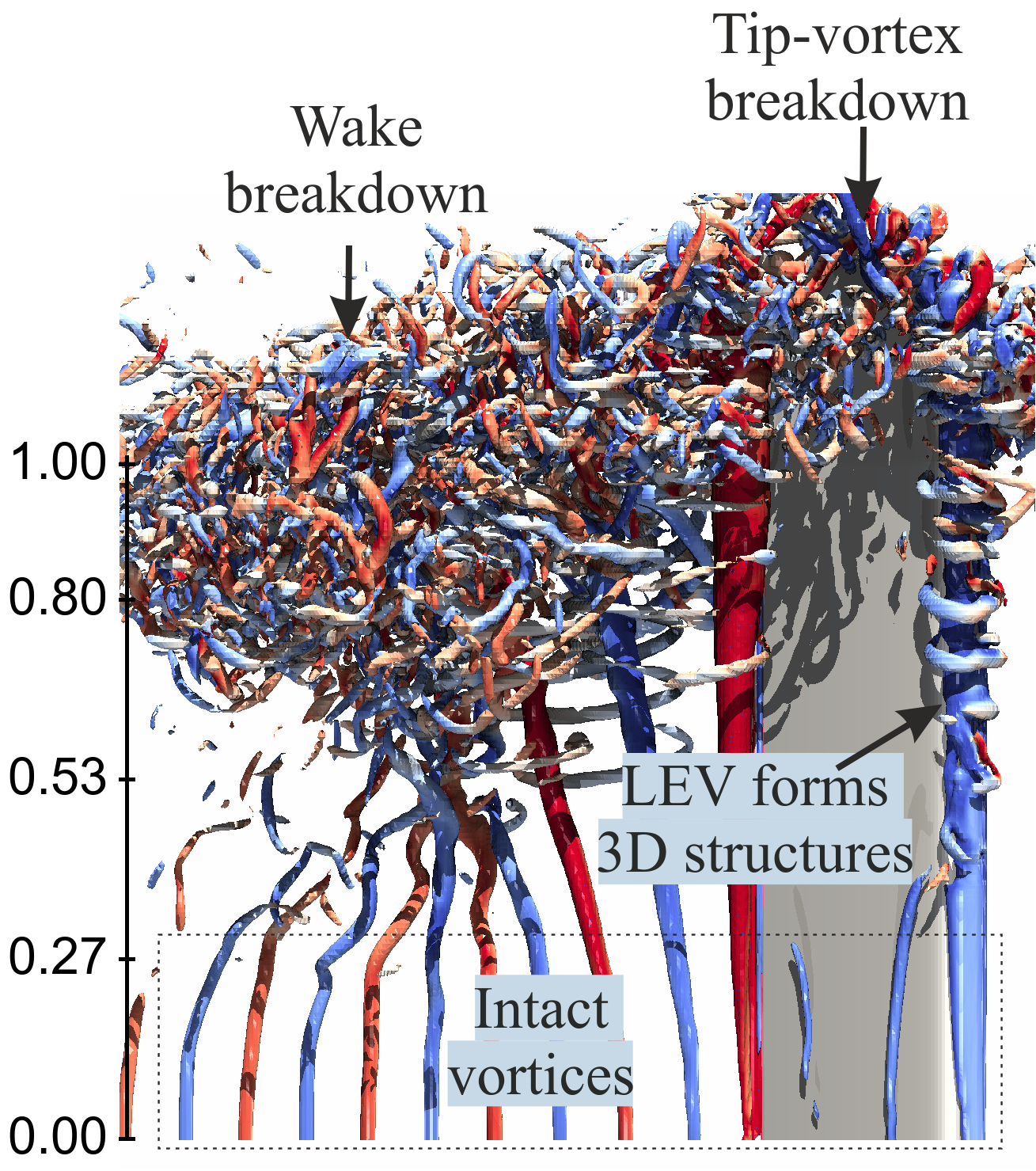} 
	& \includegraphics[trim=0 0 0 0, clip=true, width=0.25\textwidth]{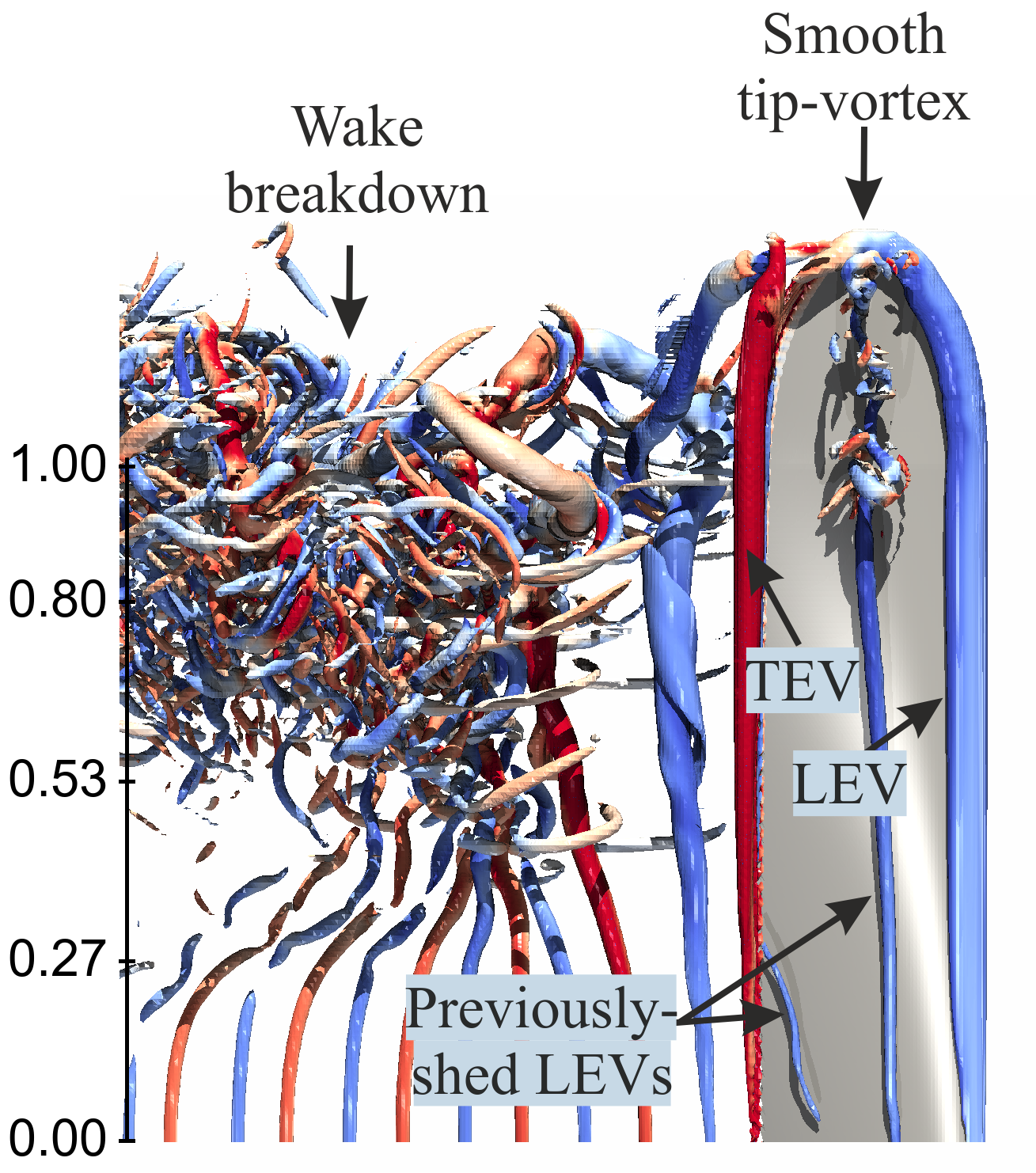}\\
	\hline
	\end{tabular}
	\includegraphics[width=0.3\textwidth]{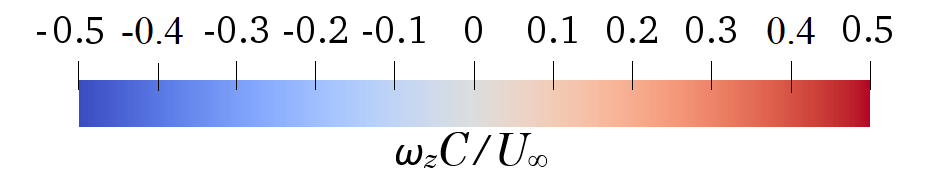} 
	\end{center}
\caption{Instantaneous vortex wake visualization at $t/T=0.5$ for both kinematics and $S=3C$. Vortices are visualized with isocontours of $\lambda_2$ colored by dimensionless spanwise vorticity $\omega_z C/U_{\infty}$. Free-stream flow $U_\infty$ is flowing right to left and spanwise scale given in $z/S$.}
\label{fig:inst}
\end{figure}

\begin{figure*}
	\begin{center}
	\begin{tabular}{|m{1.2cm}|c|c|c|c|}
	\hline
	Motion		& $t/T=0$ & $t/T=0.25$ & $t/T=0.5$ & $t/T=0.75$\\
	\hline
	Roll				& \includegraphics[trim=0 0 0 0, clip=true, width=0.2\textwidth]{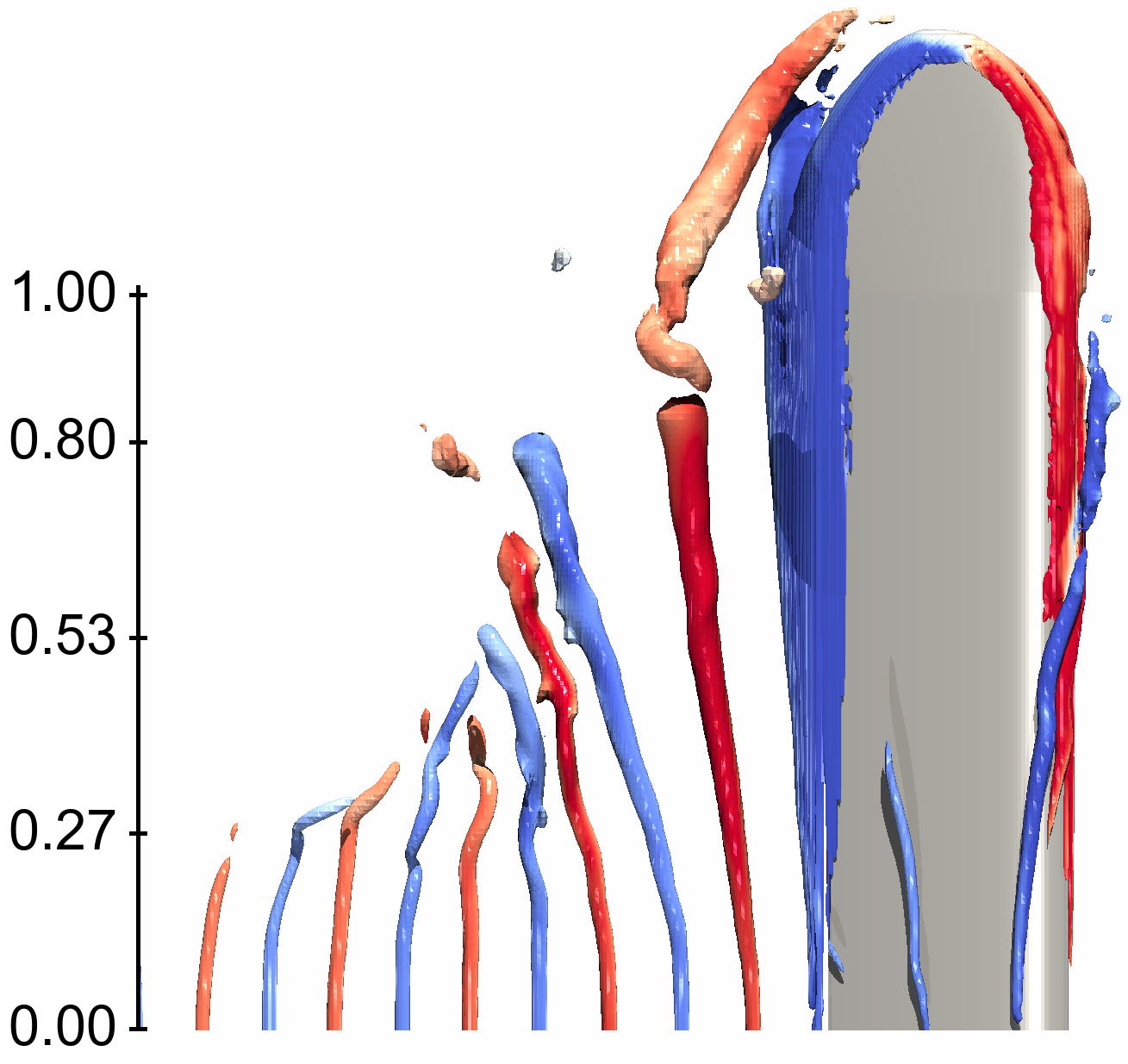} 
						& \includegraphics[trim=0 0 0 0, clip=true, width=0.2\textwidth]{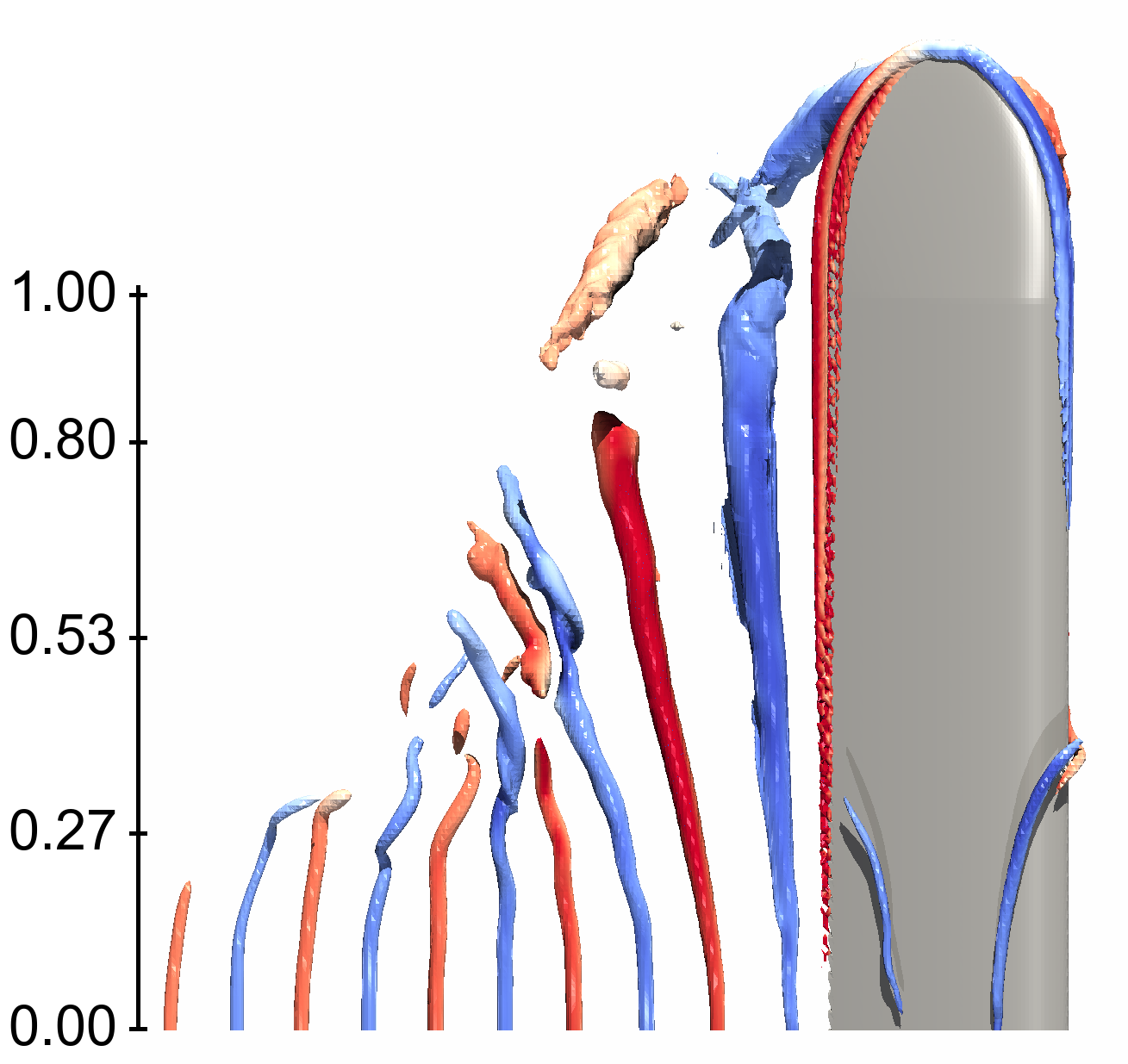}
						& \includegraphics[trim=0 0 0 0, clip=true, width=0.2\textwidth]{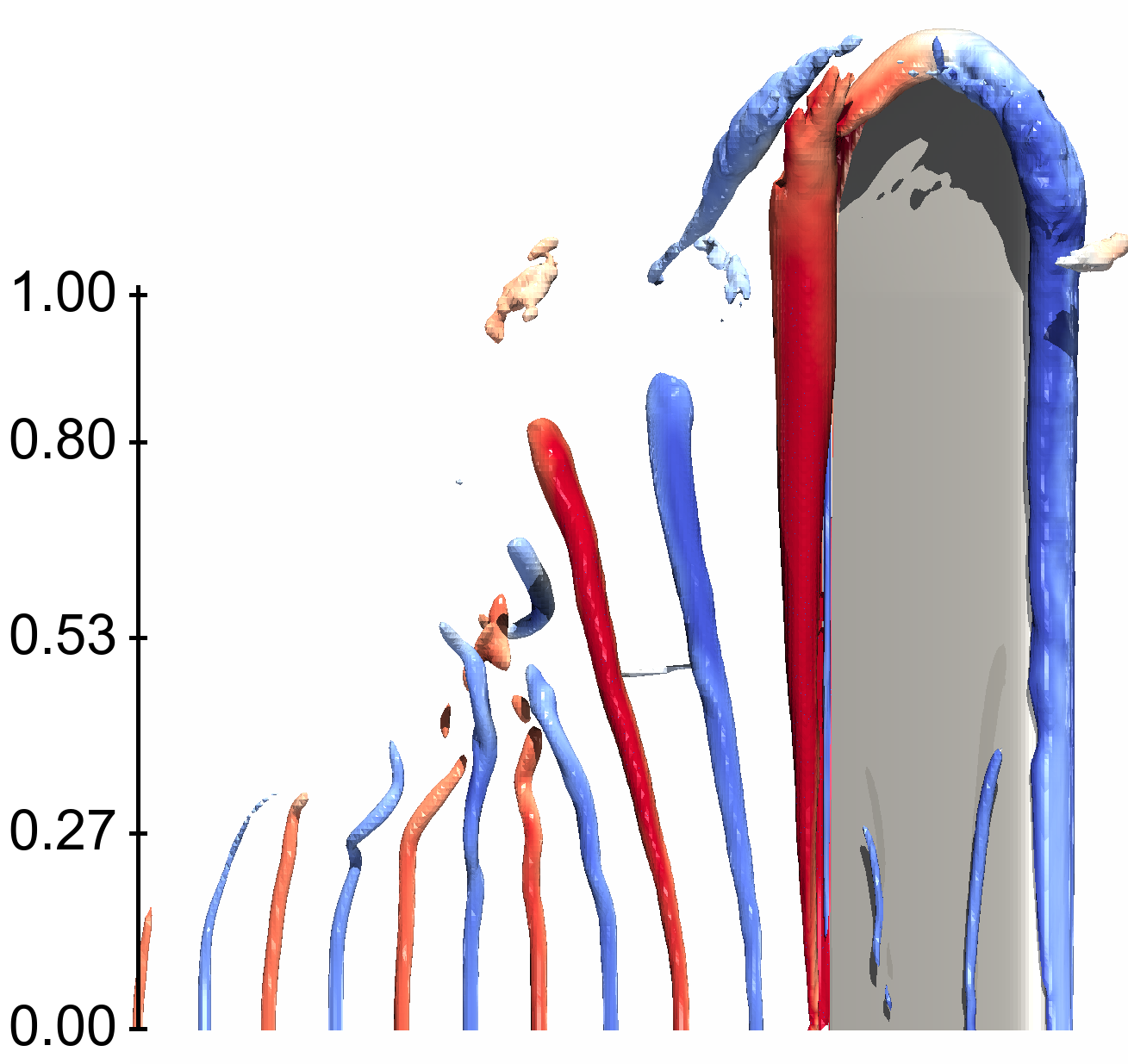}
						& \includegraphics[trim=0 0 0 0, clip=true, width=0.2\textwidth]{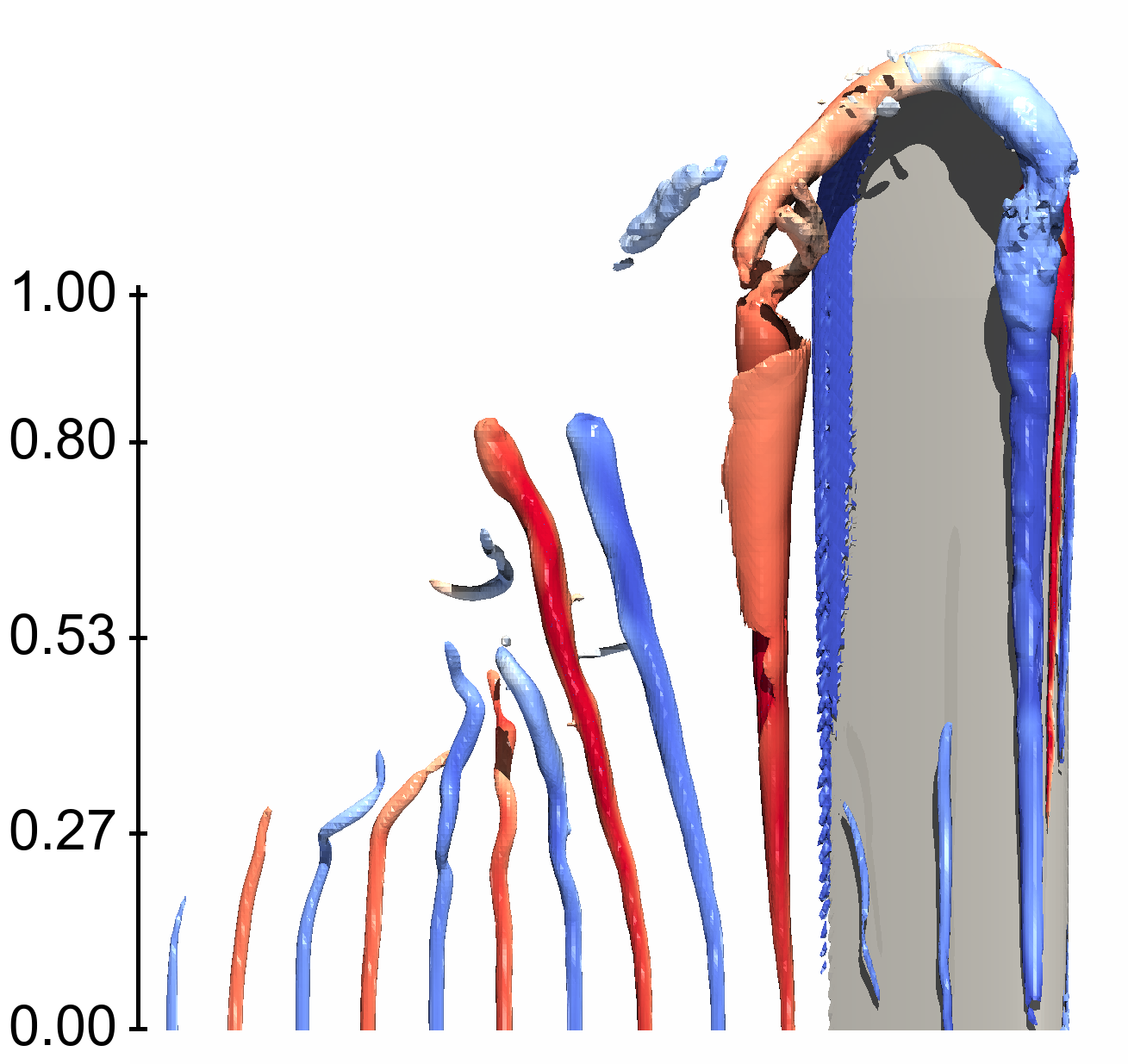}\\
	\hline
	Twist-Roll 		& \includegraphics[trim=0 0 0 0, clip=true, width=0.2\textwidth]{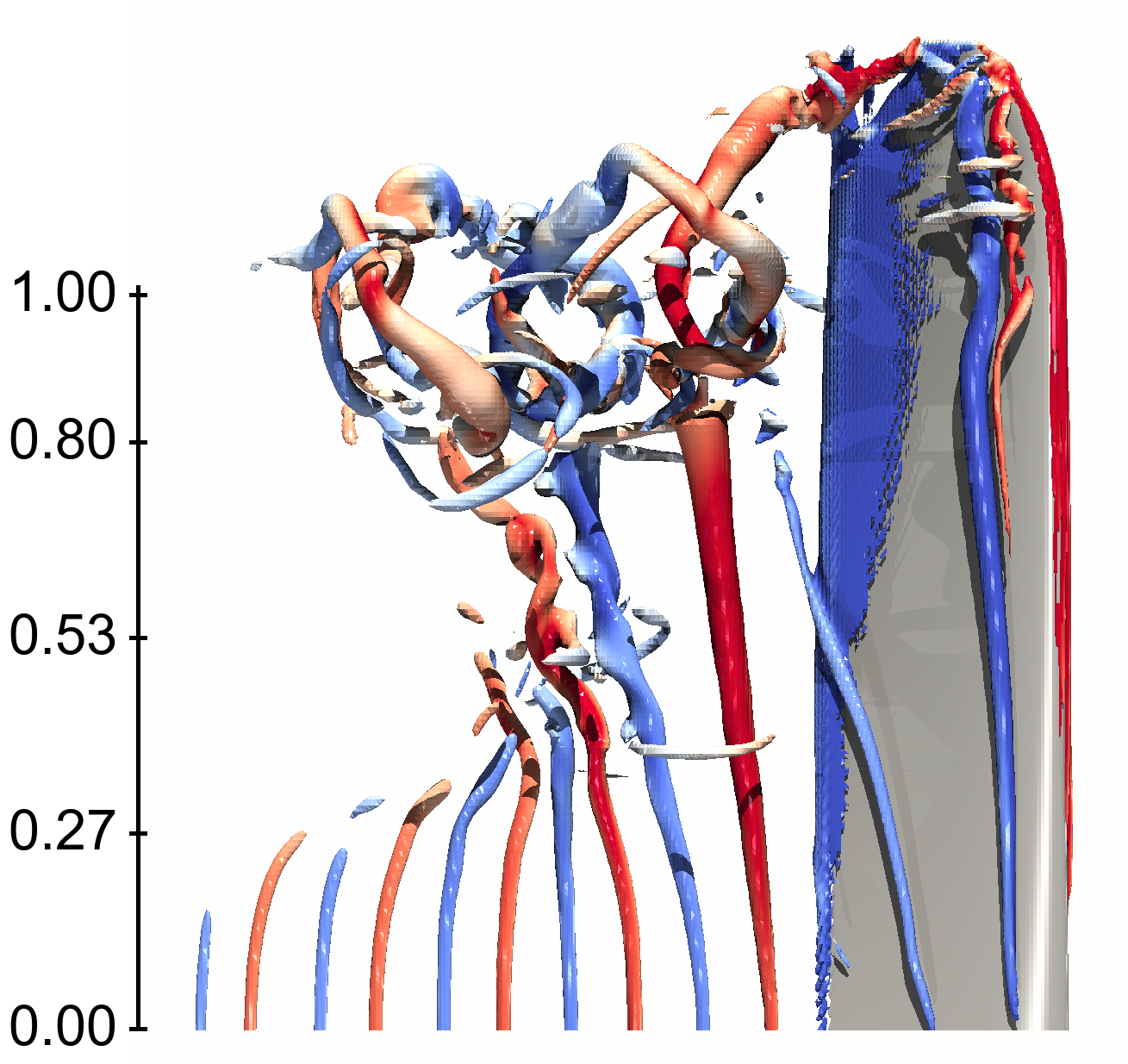} 
						& \includegraphics[trim=0 0 0 0, clip=true, width=0.2\textwidth]{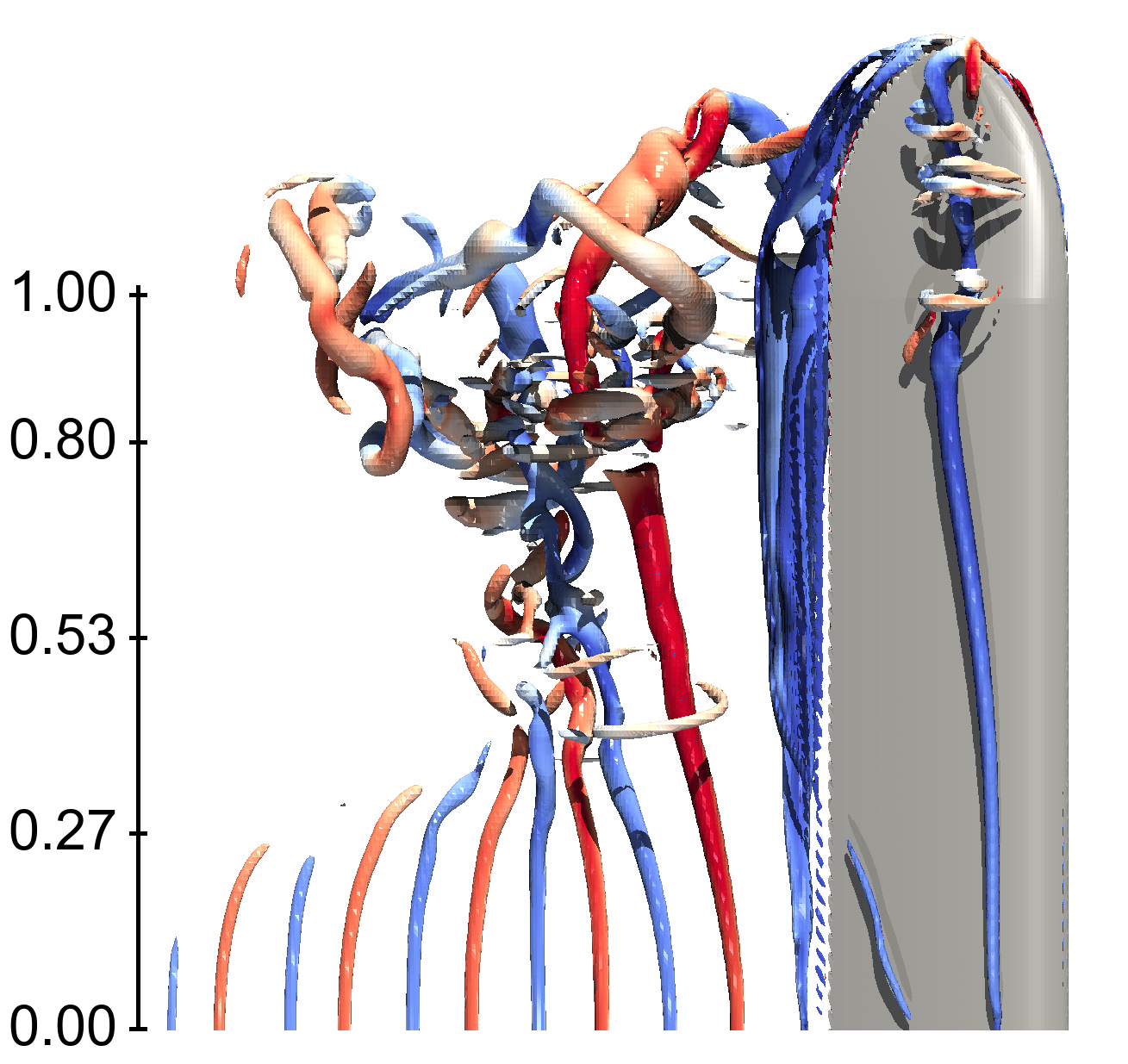}
						& \includegraphics[trim=0 0 0 0, clip=true, width=0.2\textwidth]{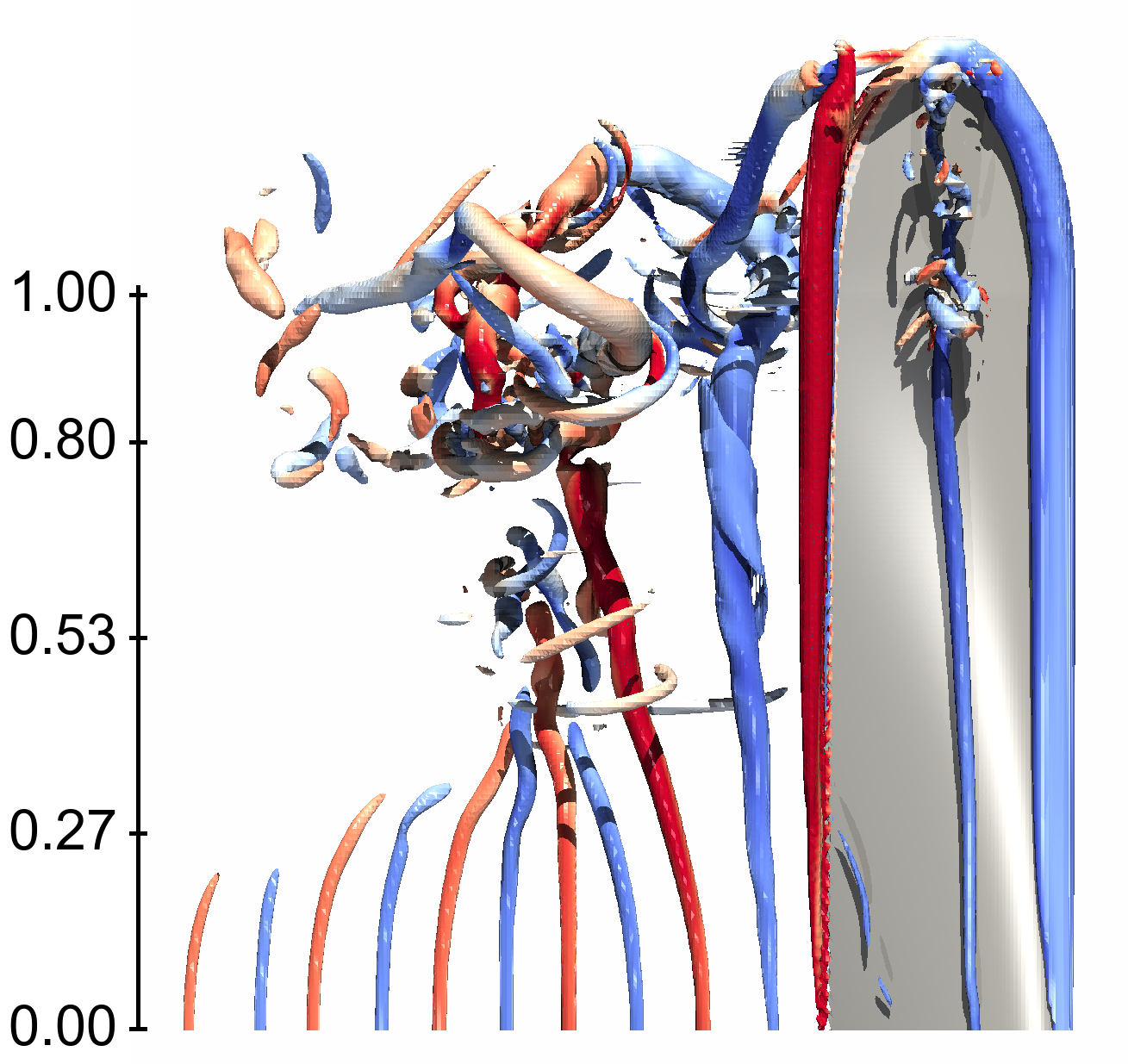} 
						& \includegraphics[trim=0 0 0 0, clip=true, width=0.2\textwidth]{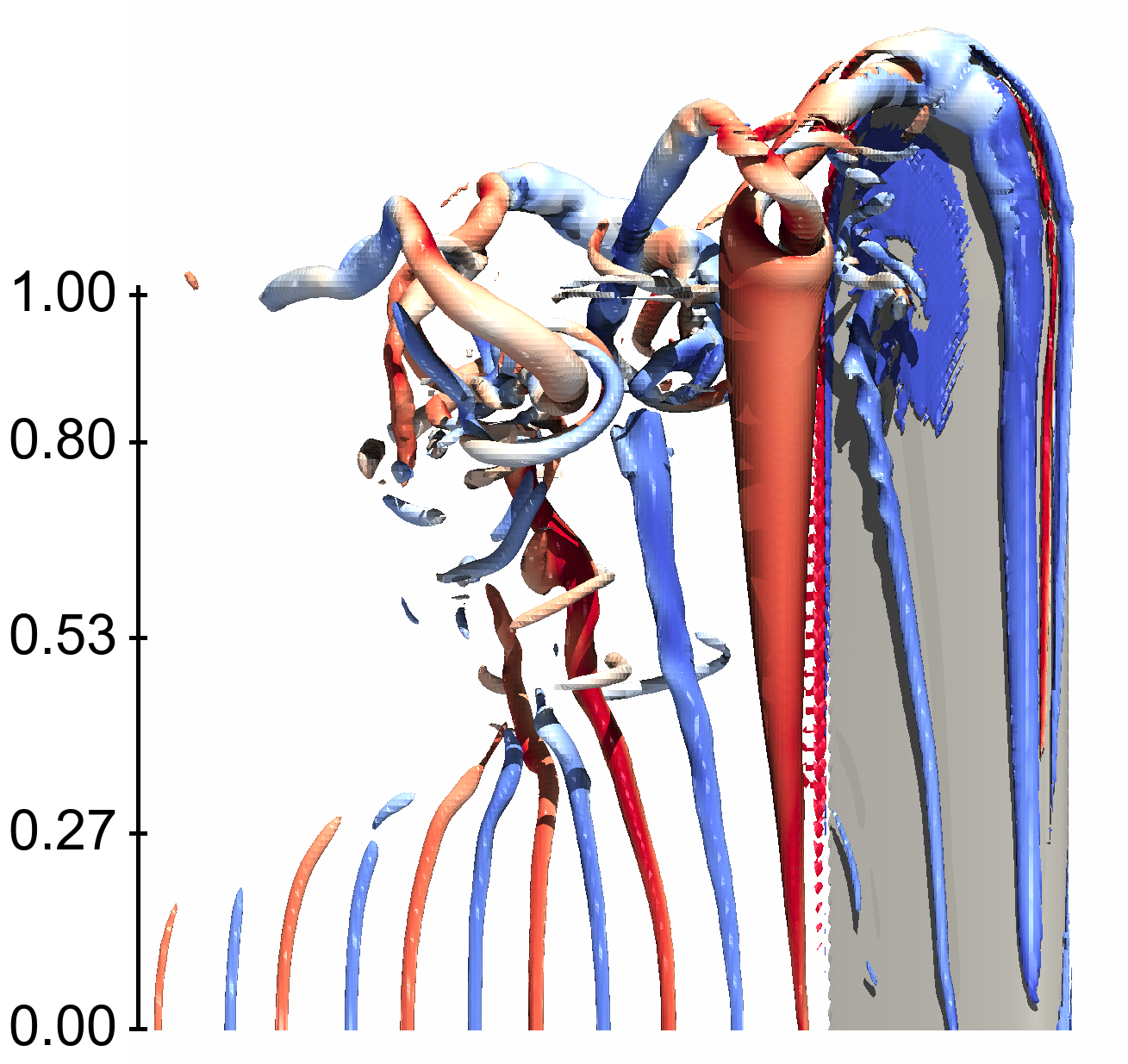}\\
	\hline
	\end{tabular}
	\includegraphics[width=0.3\textwidth]{fig1/colorbar} 
	\subcaption{}
	\end{center}
\caption{Phase-averaged velocities for pure rolling and twisting-rolling motions for $S=3C$. Visualization as in Fig.~\ref{fig:inst}.}
\label{fig:vorticity}
\end{figure*}

This section presents the qualitative features of the flow structures induced by the 3D kinematics on the finite foil. As a representative result,  Fig. \ref{fig:inst} visualizes an instantaneous snapshot of the turbulent vortex wake using $\lambda_2$ iso-contours colored by the spanwise vorticity $\omega_z C/U_{\infty}$ for an intermediate aspect ratio $S=3C$ and for both 3D kinematics. The small turbulent structures indicating vortex breakdown are present for both kinematics. In order to focus on the most important aspects of the wake, the flow was phase-averaged at phases $t/T=0,1/4,1/2,3/4$. Fig. \ref{fig:vorticity} shows this simplifies the visualization, revealing the dominant structures which occur every cycle and removing higher frequency turbulent structures related to vortex breakdown. The pure roll and twist-roll cases have distinct vortex structures due to their different kinematics, but the periodic formation of the vortices such as the leading-edge (LEVs), trailing-edge (TEVs) and tip vortices are much more clear in the phase-averaged representation. 

Both the instantaneous flow (Fig. \ref{fig:inst}) and the phase-averaged flow (Fig. \ref{fig:vorticity}) show strong three-dimensionality in their vortex structures along the span. Because the amplitude of motion increases linearly along the span, the sectional Strouhal $St_A$ increases linearly as well. Starting from the root, we see the vortices are initially coherent and skewed obliquely, tilted further downstream as you progress towards the tip. This is especially clear in the near phase-averaged wake, where coherent structures persist all the way to the tip. Both the coherence and skew are in direct contrast to the complete 3D vortex breakdown found in the wake of infinite foils heaving and pitching with the same amplitudes \cite{zurman2020}, and therefore this must be due to the spanwise derivative of the kinematics not present in the infinite foil case. Indeed, the oblique near-wakes resembles the vortices of delta or swept wing and hints that the spanwise derivative of the kinematics induces a stabilizing spanwise flow \textemdash as found in swept wings and rotating foils at low revolution angles \cite{Jardin2017}. For this finite foil, the far wake vortex structures break down into 3D non-periodic turbulence above section $z/S=0.27$. This corresponds to a sectional $St_A=0.3$ where 2D uniform structures were observed for both pure heaving and heaving-pitching combination for the infinite foil cases in \citet{zurman2020}. 

Contrasting the pure roll to the twist-roll case, we see that the tip vortex flow is much cleaner in the twist-roll case. The correspondence between the instantaneous and phase-averaged plots shows that the LEV vortex is essentially periodic near the foil, and does not breakdown until further downstream in the wake. This is due to the twist motion being out of phase with roll, reducing the angle of attack on the tip, Fig \ref{fig:aoa}, and thereby reducing the strength of the tip vortices. This is further quantified in the next section. In contrast, the pure rolling TEV begins to breakdown as soon as it separates, see Fig. \ref{fig:inst} and supplementary videos. This also implies that the foil tip is more lightly loaded in the roll-twist case, reducing the lift forces that can be achieved at compared to rolling kinematics, as will be discussed further in the following sections.

\subsection{Aspect-ratio effects}
\label{aspect_ratio}

\begin{figure}
	\begin{center}
	\begin{tabular}{|c|b{4.7cm}|}
	\hline
	3C		&   \multicolumn{1}{c|}{6C} \\
	\hline
	\includegraphics[trim=0 0 0 0, clip=true, width=0.2\textwidth]{fig1/roll3c_fl_top3} 
	& \includegraphics[trim=0.5cm 0 0.5cm 0, clip=true, width=0.26\textwidth]{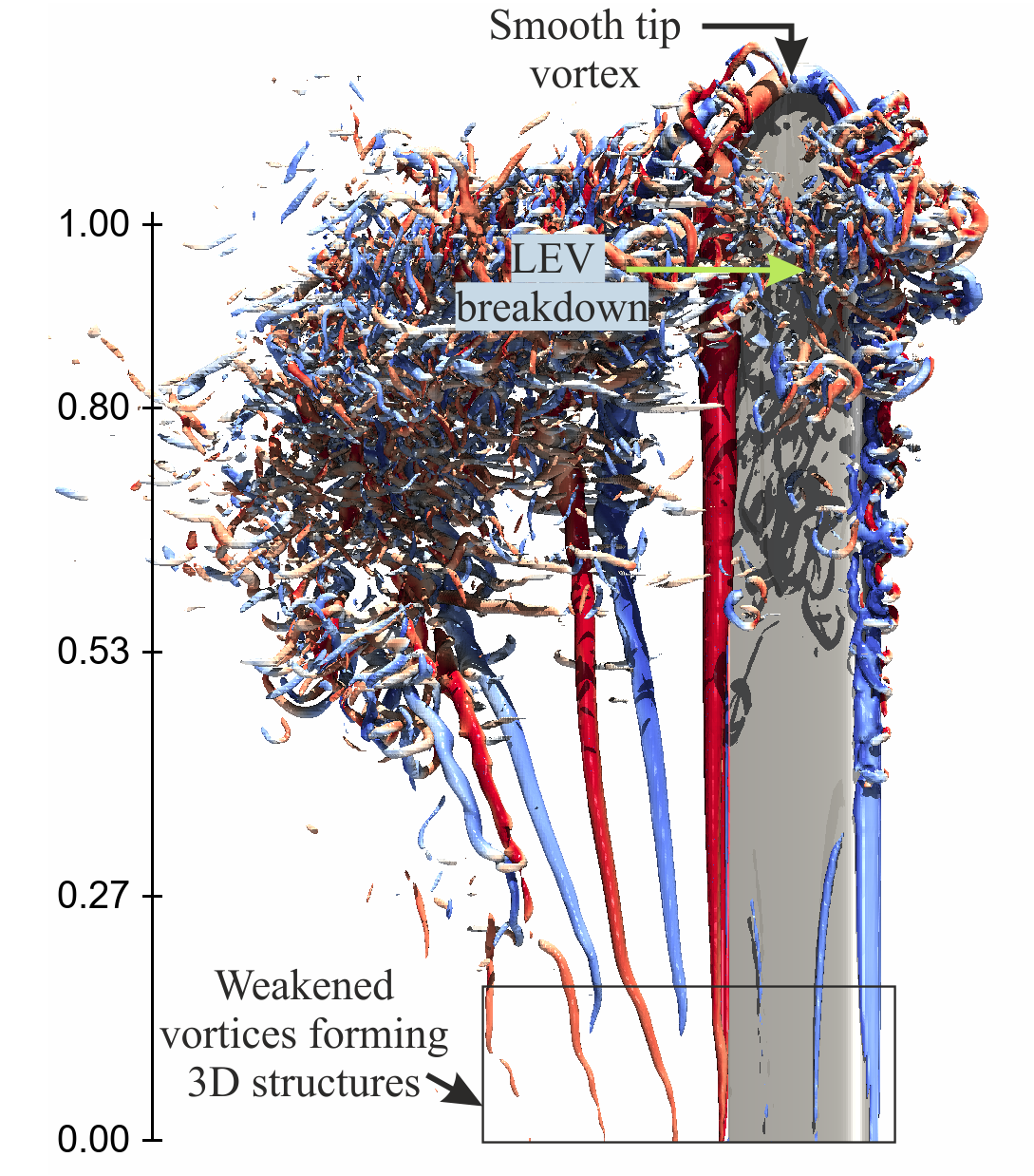}\\
	\hline
	\end{tabular}
	\includegraphics[width=0.3\textwidth]{fig1/colorbar} 
	\end{center}
\caption{Instantaneous flow at $t/T=0.5$ for finite foils undergoing the same pure roll kinematics. Labels contrast the relatively weaker wake vortices on the centerline ($z=0$) and less turbulent tip flow on the $S=6C$ foil compared to $S=3C$. Visualization as in Fig.~\ref{fig:inst}.}
\label{fig:inst_roll}
\end{figure}

This section presents results on the influence of \AR\xspace on the flow and forces. Fig. \ref{fig:inst_roll} compares the instantaneous wake structures for $S=3C$ and $S=6C$ for a foil undergoing pure roll. Note that the 2D kinematics on each $z/S$ slice match between these cases, e.g. $St_A=0.3$ on $z/S=0.27$ for both foils. Therefore the strip theory solution is identical for the two cases, while the 3D simulations show obvious important differences. Because the span is twice as long, the spanwise derivatives of the kinematic parameters $St_A,\alpha,\dot\alpha$ are half as strong, weakening the three-dimensionality of the flow. First, the turbulent tip flow and wake breakdown is less uniform in the $S=6C$ case, with a few clean vortex structures identifiable over a longer span length. In addition, the center-line ($z=0$) reversed K\'arm\'an vortex street is completely different between the $3C$ and $6C$ cases. As the foil is motionless at the centerline, these vortices would not even appear in a strip theory approach, and as they are far from the tip, they isolate the influence of the spanwise derivative of the kinematics on the 3D flow. Fig. \ref{fig:inst_roll} shows that the $S=6C$ foil allows characteristics of the infinite foil to appear, including very weak and 3D vortex structures in regions with very low amplitude \cite{zurman2020}. This area is dominated by the stationary foil vortex-shedding induced only by the bias angle of attack $\theta_{bias}=10^\circ$.

\begin{figure*}
	\begin{center}
	\begin{tabular}{|c|c|c|c|b{4cm}|}
	\hline
	Motion		& $1C$ & $2C$ & $3C$ & \multicolumn{1}{c|}{$6C$}\\
	\hline
	Roll			& \includegraphics[trim=0 0 0 0, clip=true, width=0.2\textwidth]{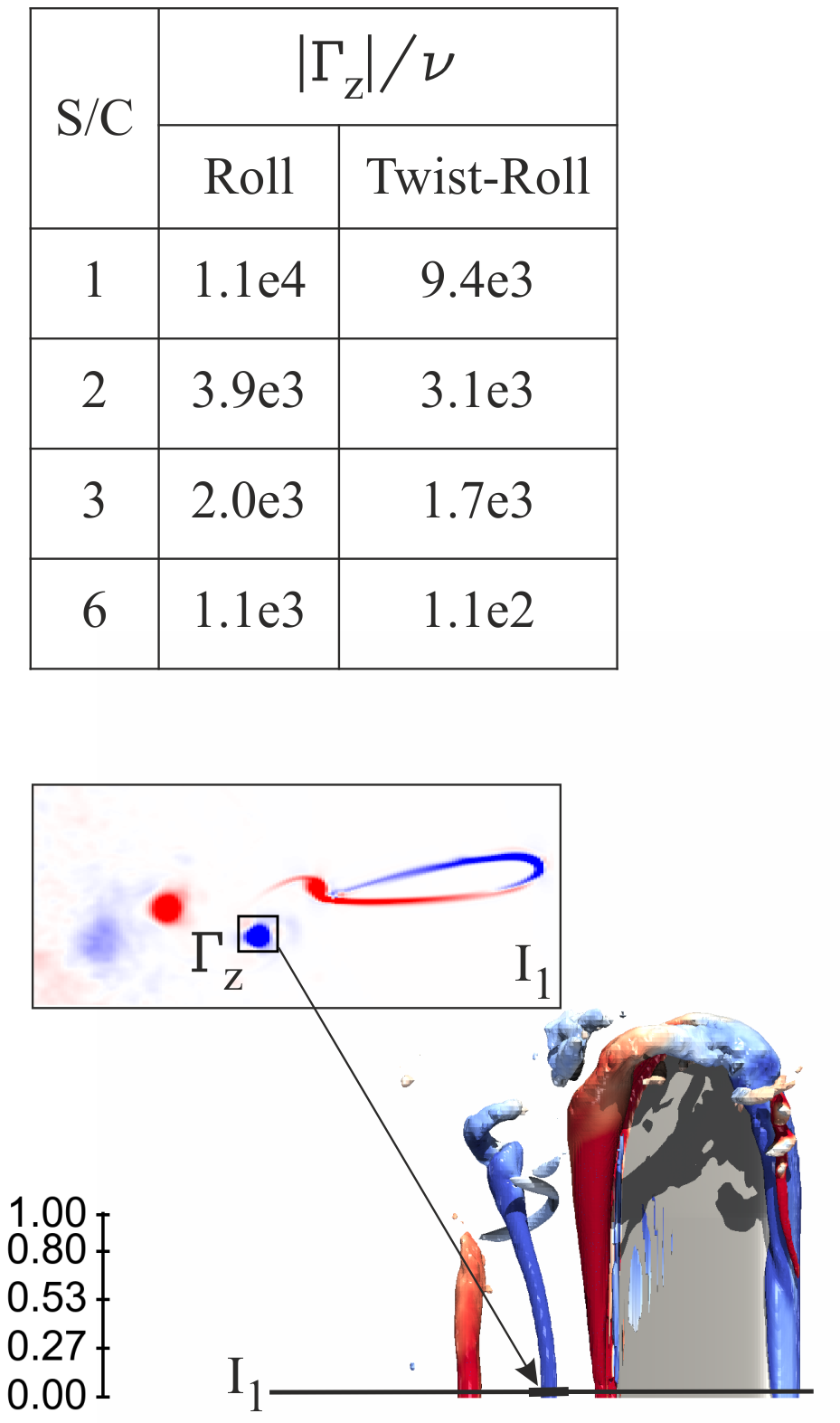} 
						& \includegraphics[trim=0 0 0 0, clip=true, width=0.2\textwidth]{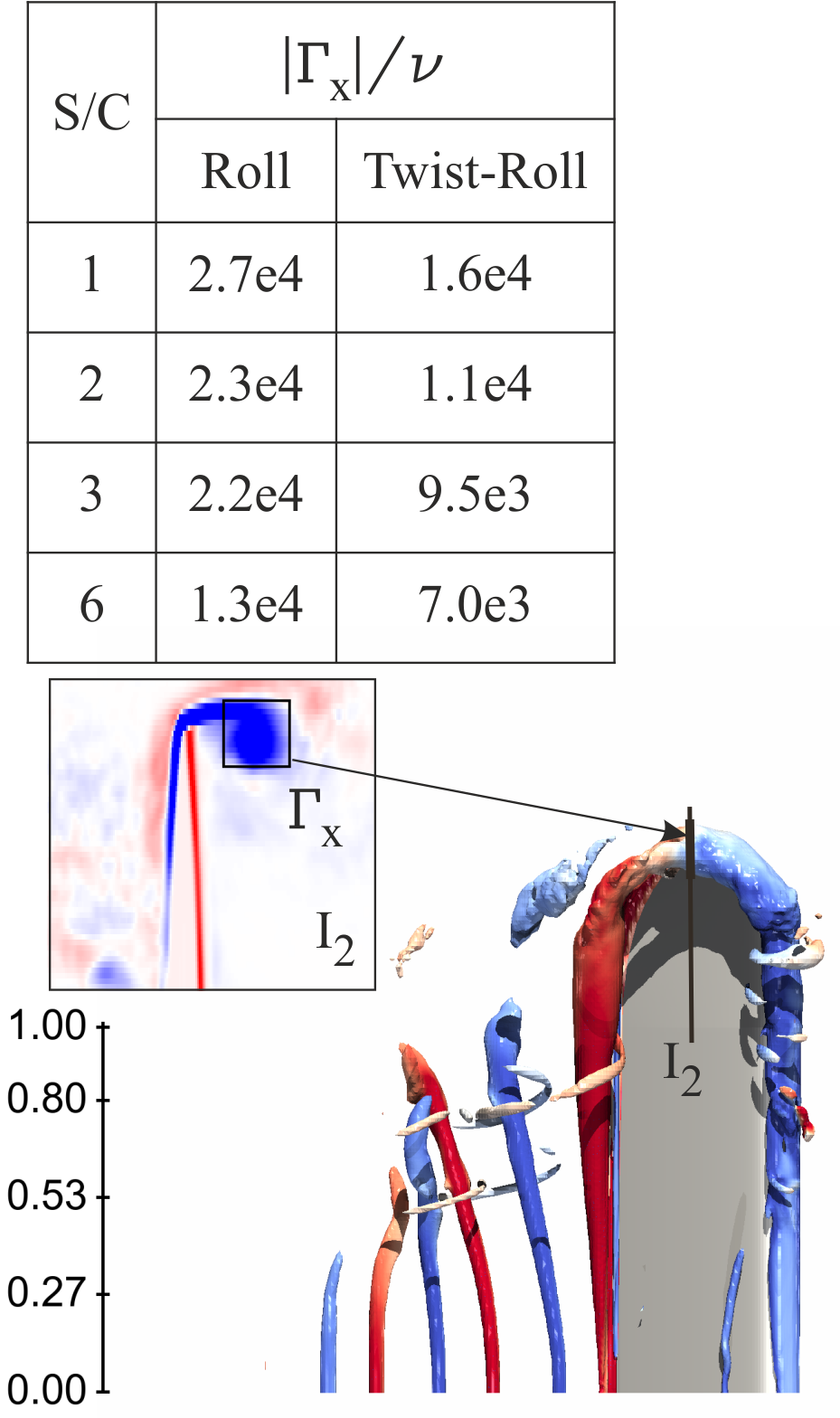}
						& \includegraphics[trim=0 0 0 0, clip=true, width=0.2\textwidth]{fig1/roll3c_top3}
						& \includegraphics[trim=0 0 0 0, clip=true, width=0.26\textwidth]{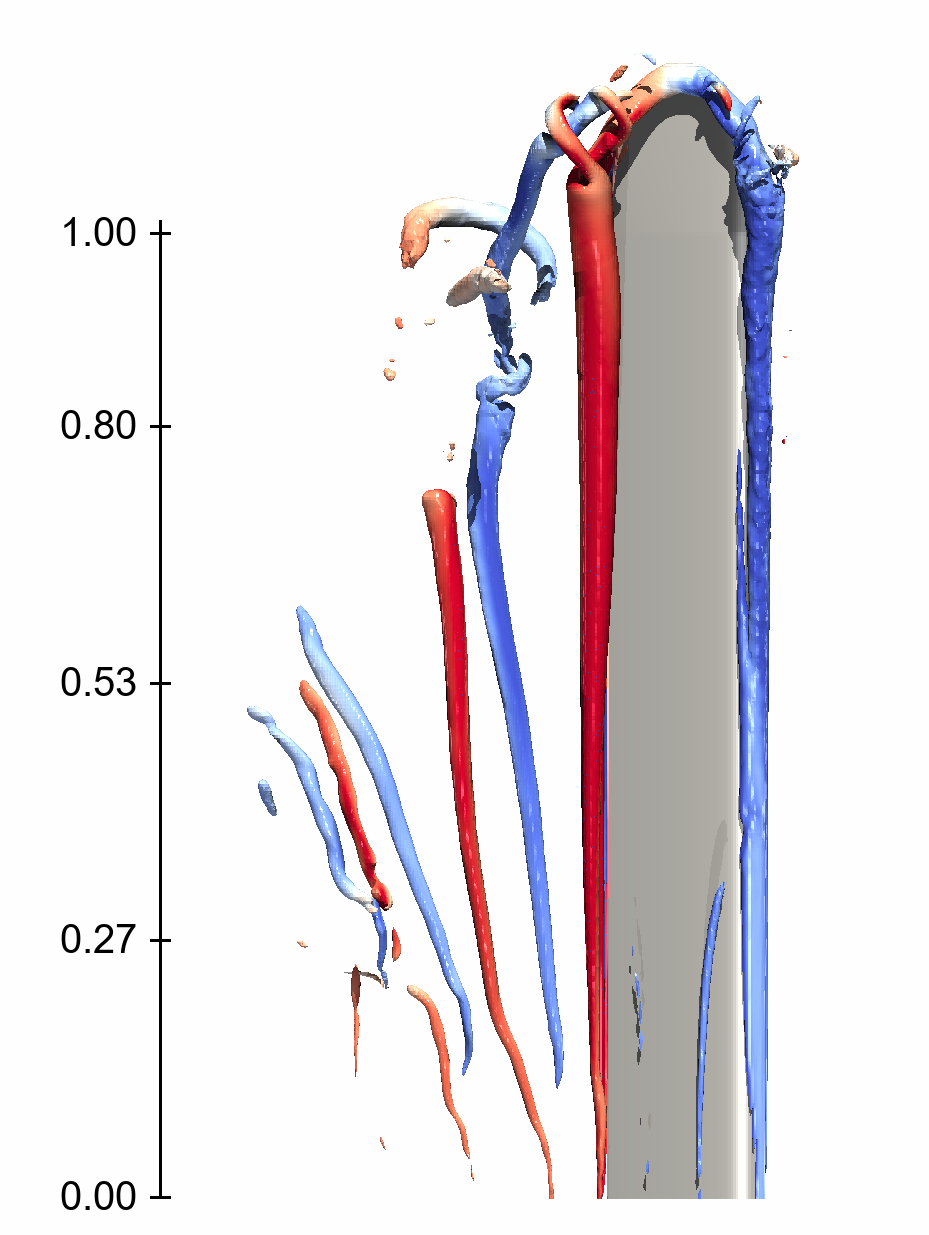}\\
	\hline
	Twist-roll 	& \includegraphics[trim=0 0 0 0, clip=true, width=0.2\textwidth]{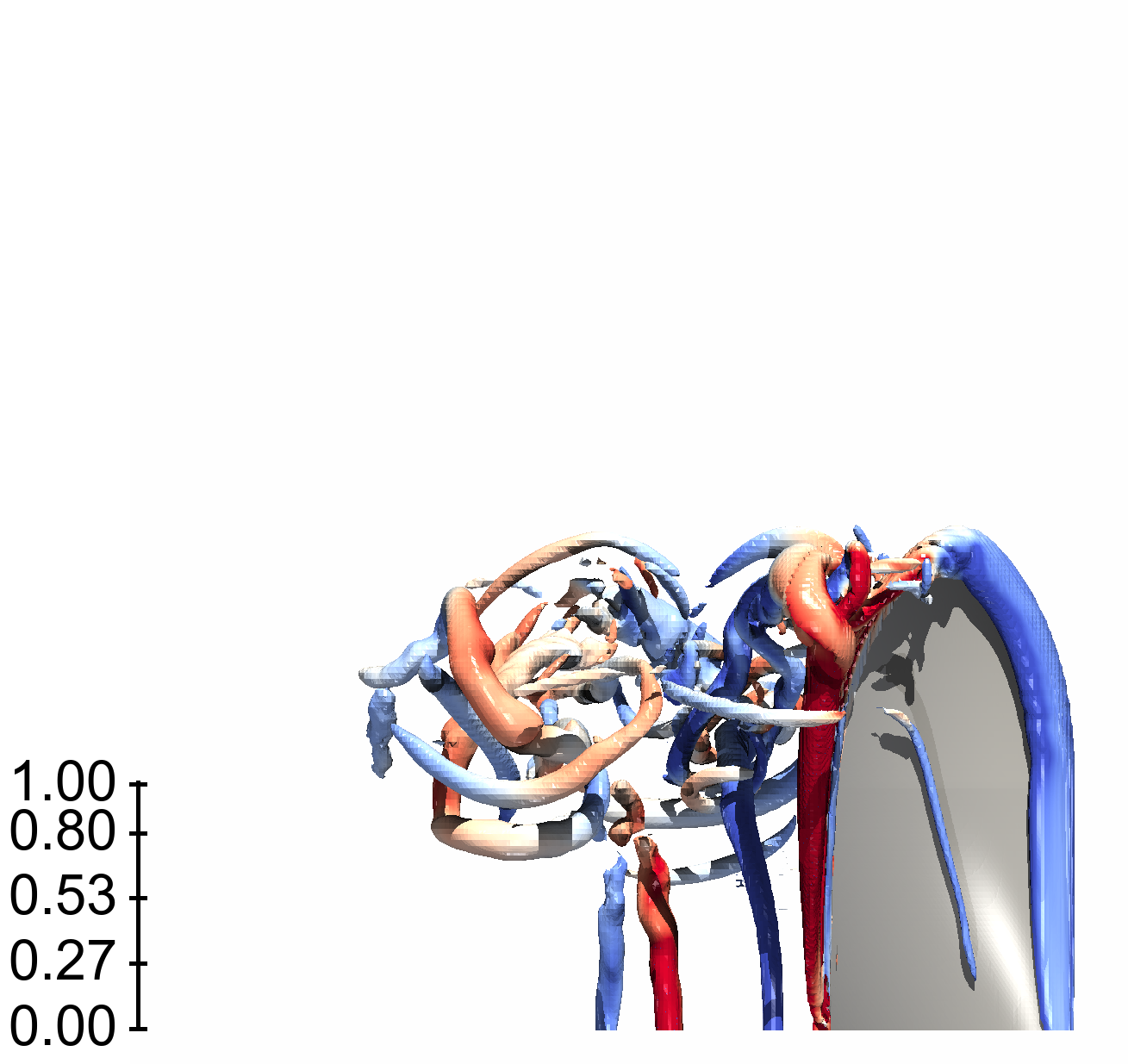} 
						& \includegraphics[trim=0 0 0 0, clip=true, width=0.2\textwidth]{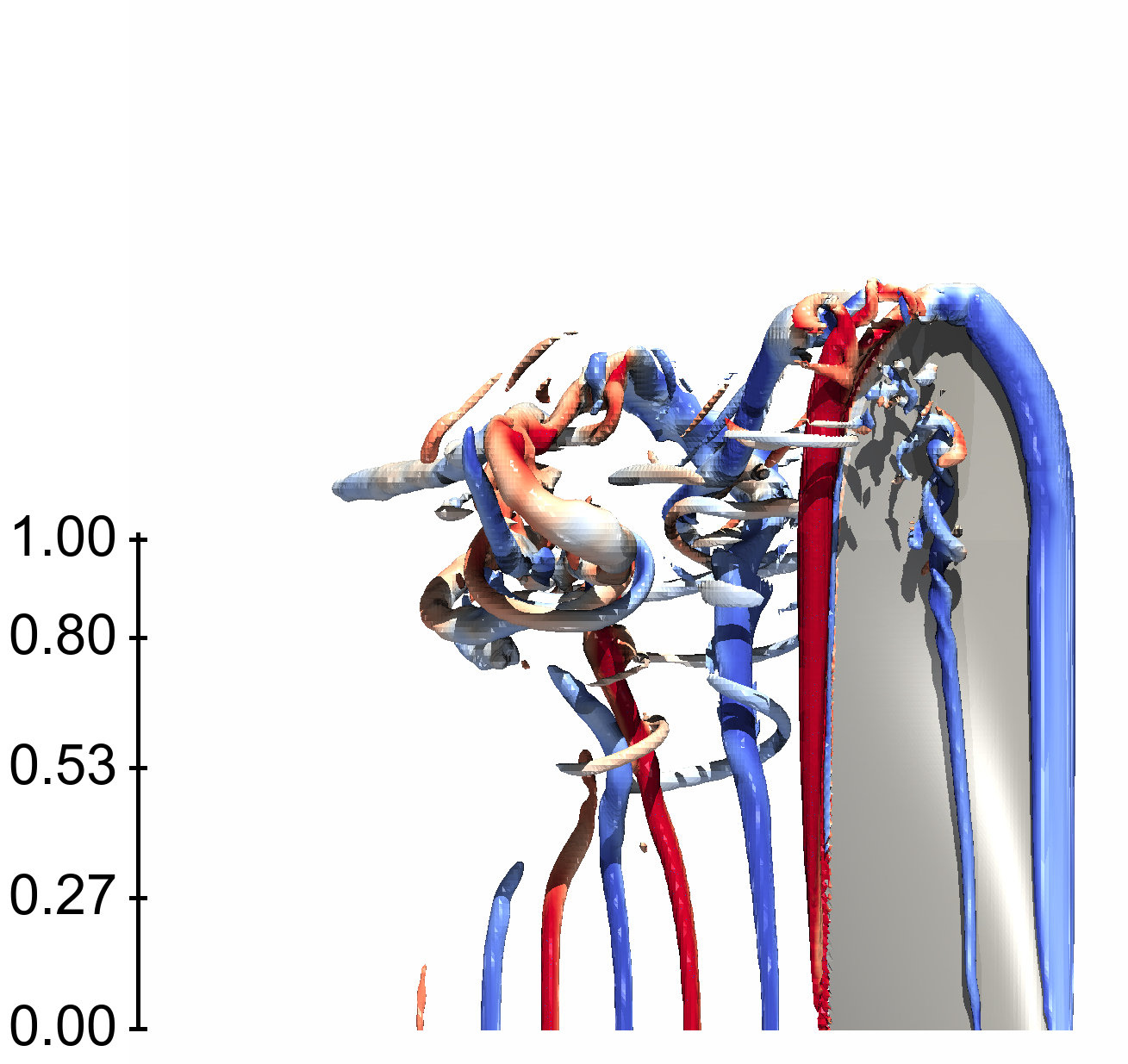}
						& \includegraphics[trim=0 0 0 0, clip=true, width=0.2\textwidth]{fig1/twistroll3c_top3} 
						& \includegraphics[trim=0 0 0 0, clip=true, width=0.26\textwidth]{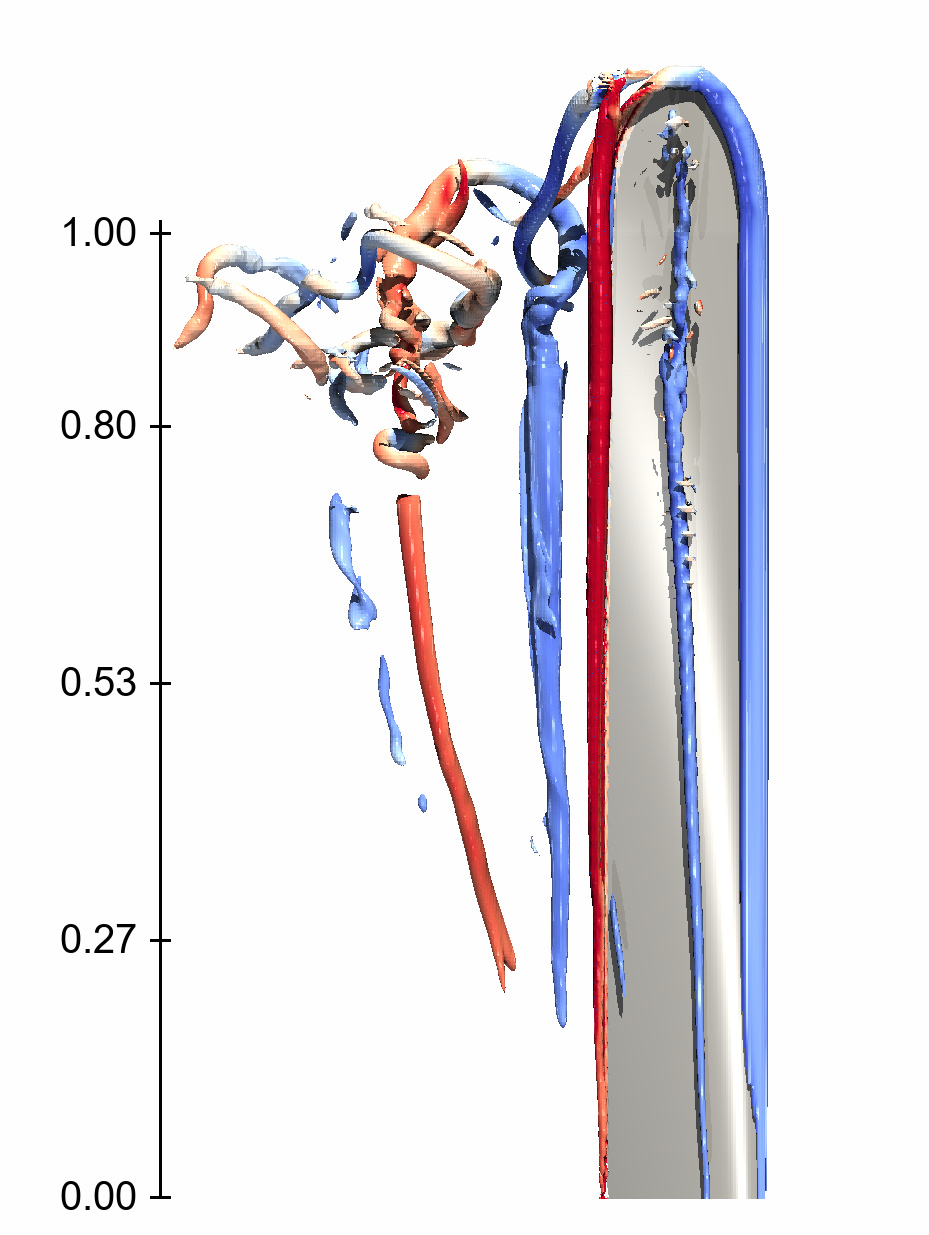}\\
	\hline
	\end{tabular}
	\includegraphics[width=0.3\textwidth]{fig1/colorbar} 
	\end{center}
\caption{Phase-averaged flow for pure roll and twist-roll motions for different aspect ratios at $t/T=0.5$. Flow visualization as in Fig.~\ref{fig:inst}. Inset figure $I_1$ shows the spanwise vorticity on a center-line slice and $I_2$ shows the streamwise vorticity on a slice through the tip. The inset tables give the viscosity-scaled circulation of the shed vortices.}
\label{fig:ar_top}
\end{figure*}

These effects are reiterated and quantified in Fig.\ref{fig:ar_top} which shows the phase-averaged flow across the full range of tested $S/C$. As before, the kinematics for each $z/S$ slice match across the foil sizes. The plot also quantifies the viscosity scaled circulation of the tip vortices ($\Gamma_x$) and centerline vortices ($\Gamma_z$). The circulation is computed by integrating the phase-averaged vorticity over a rectangular slice through the vortex cores (see Fig.\ref{fig:ar_top} insets) and are kept the same size as span is varied, but not between $\Gamma_z$ and $\Gamma_x$. The tip vortices are seen to dominate a larger portion of the flow as \AR\xspace is reduced, as expected. However, the relative strength of the tip vortices only grows by a factor of 2 as the span is reduced from $S=6C$ to $S=C$ for either motion type. A much stronger effect is observed in the center-line ($z=0$) circulation of the vortices in the reversed K\'arm\'an street.  Fig.\ref{fig:ar_top} quantifies that these vortices strengthen by an order of magnitude when the span is reduced from $S=6C$ to $S=C$ for the pure roll case, and nearly two orders of magnitude for the twist-roll case. 

\begin{figure*}
	\centering
	\begin{subfigure}[b]{0.34\textwidth}
		\includegraphics[trim=0 0 1.8cm 2.3cm, clip=true, width=1\textwidth]{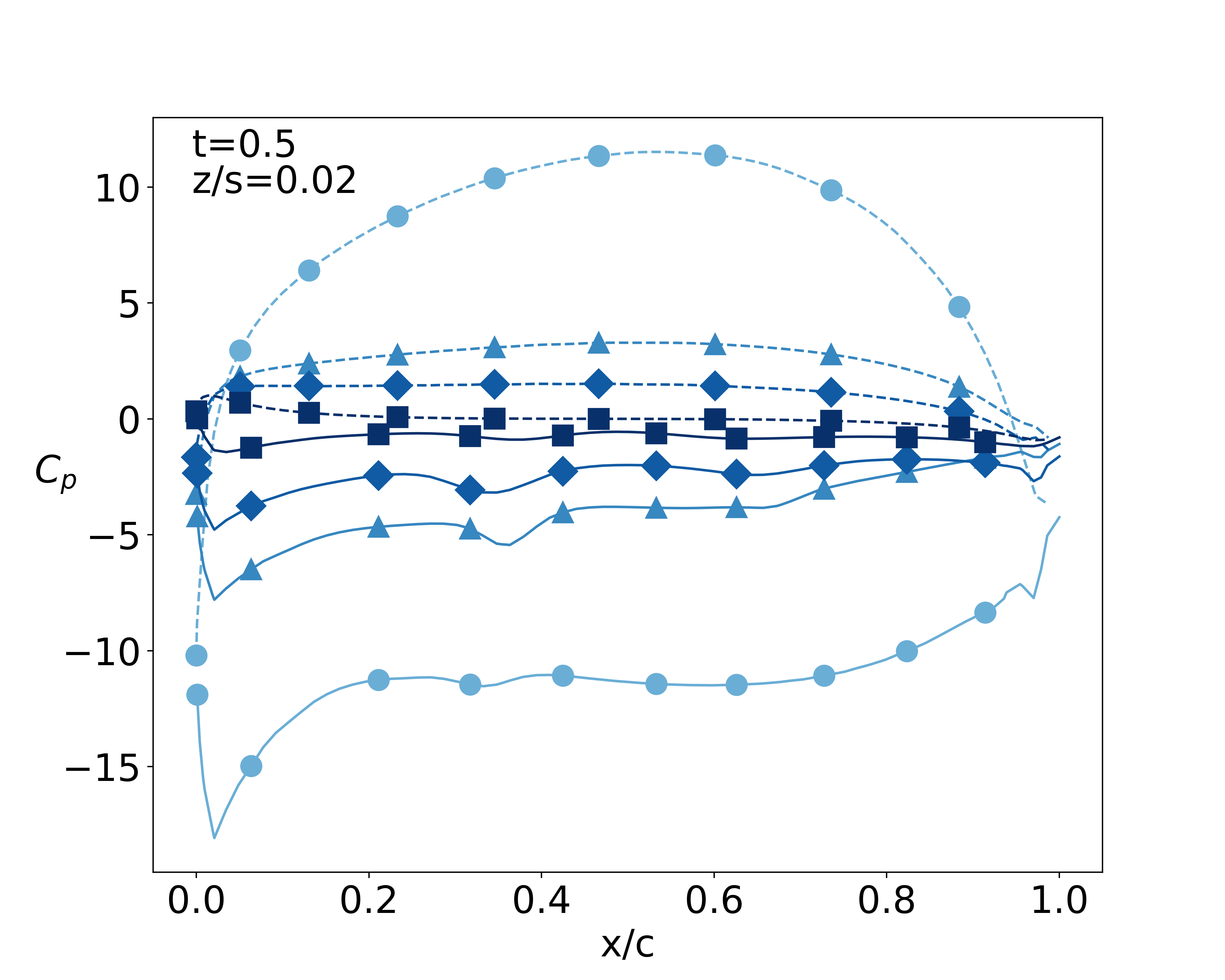}
		\subcaption{}
	\end{subfigure}
	\begin{subfigure}[b]{0.32\textwidth}
		\includegraphics[trim=1.2cm 0 2cm 2.3cm, clip=true, width=1\textwidth]{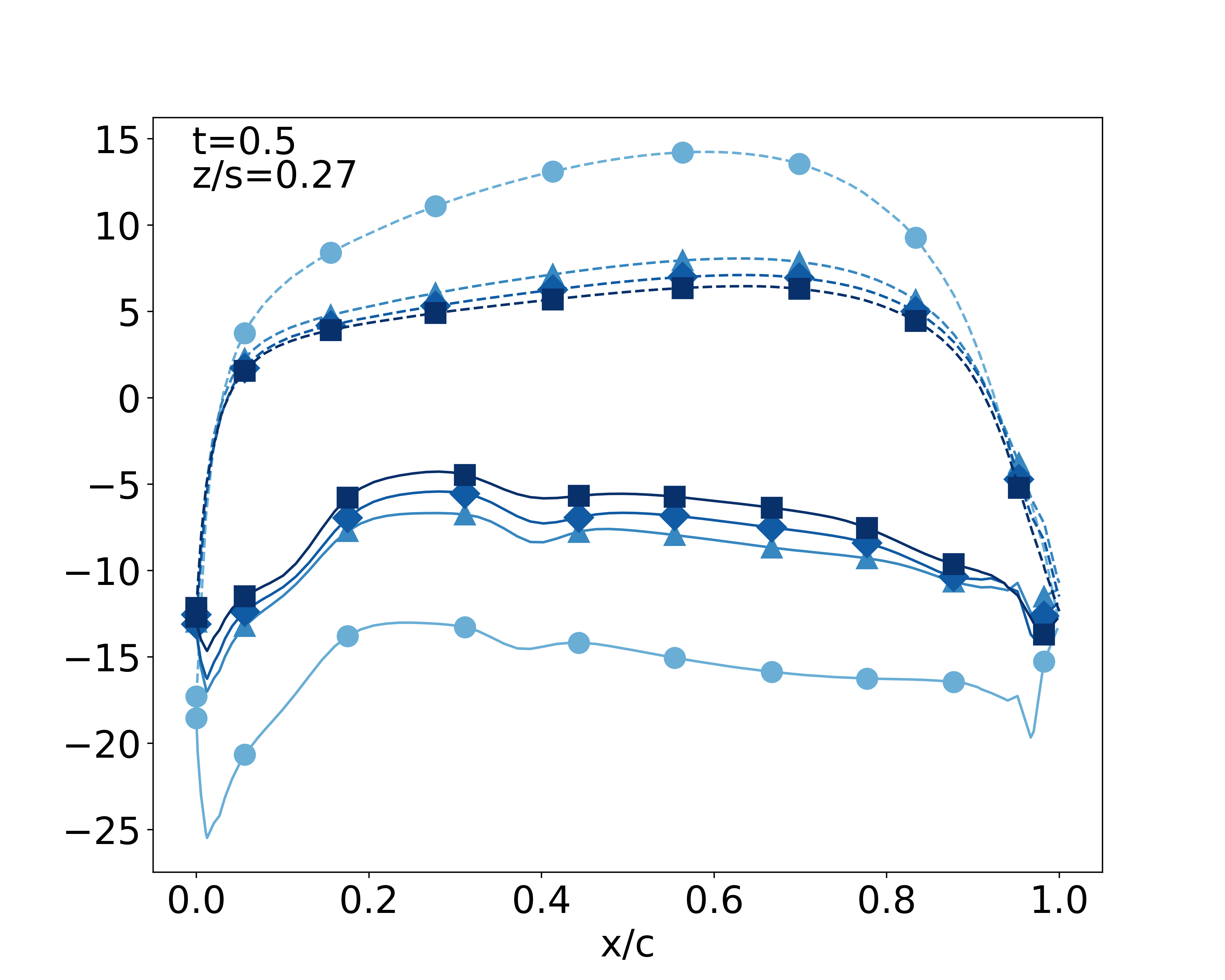}
		\subcaption{}
	\end{subfigure}
	\begin{subfigure}[b]{0.32\textwidth}
		\includegraphics[trim=1.2cm 0 2cm 2.3cm, clip=true, width=1\textwidth]{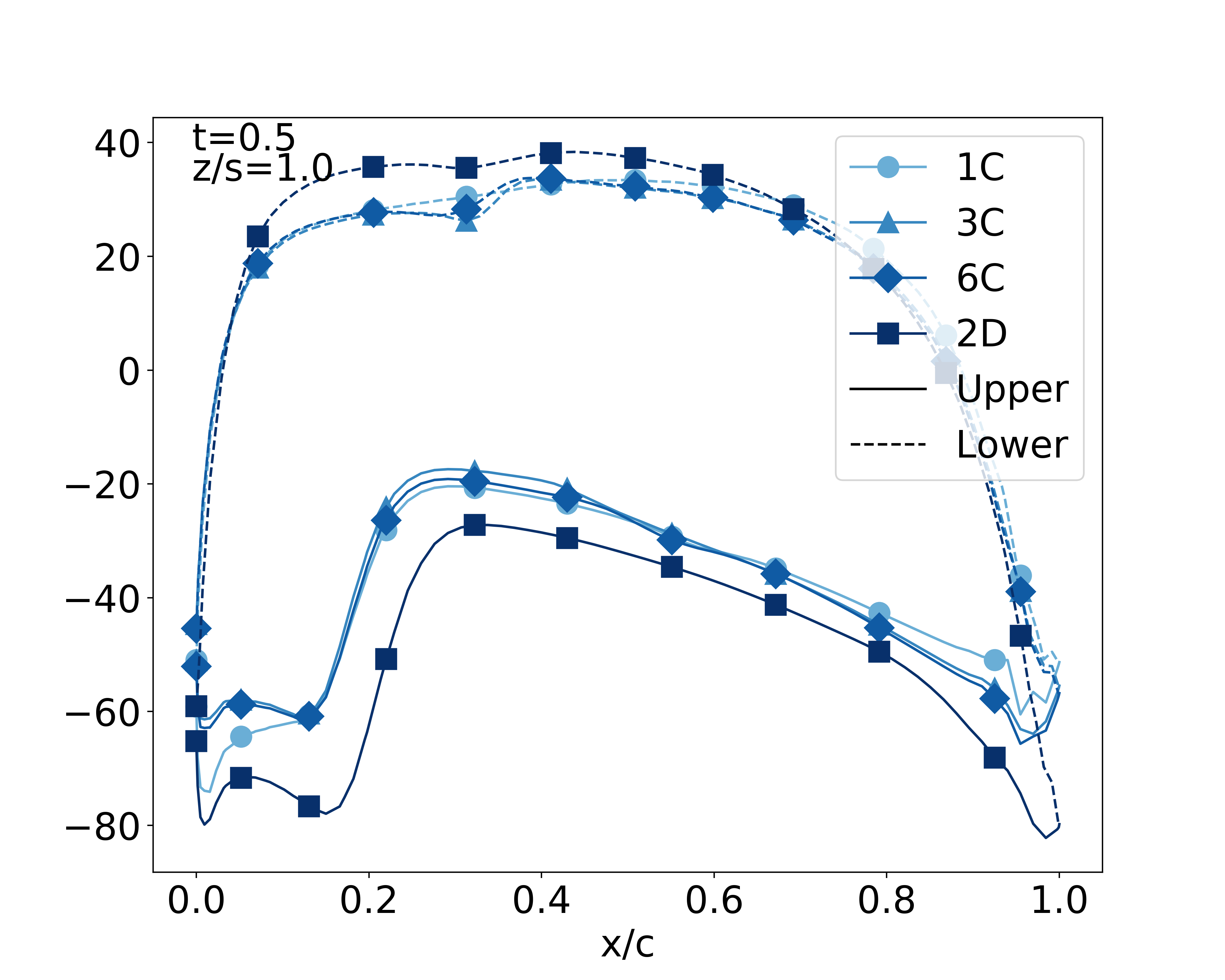}
		\subcaption{}
	\end{subfigure}
\caption{Phase-averaged pressure coefficient of twisting-rolling foil for cross sections $z/S=$ (a) $0.02$, (b) $0.27$ and (c) $1.0$, at $t/T=0.5$ for upper and lower side of the foil. Section close to tip is kinematic dominated, whereas the one close to root is gradient dominated.}
\label{fig:pressure}
\end{figure*}

To further quantify the impact of \AR\xspace on the flow, Fig.\ref{fig:pressure} shows the phase-averaged pressure coefficient $C_p=P/(0.5\rho U_\infty^2)$ at three slices along the span for the twist-roll case. The roll case is similar and so not shown. We find that all \AR\xspace produce a similar $C_p$ distribution near the tip ($z/S=1$), whereas the near-root pressure coefficient is substantially higher for lower aspect ratios. This is in agreement with the wake circulation findings and supports that the near-root sections are dominated by the spanwise derivative in the kinematics. 

\begin{figure}
	\begin{subfigure}[hb]{.222\textwidth}
		\includegraphics[trim=0 0 2.cm 1.8cm, clip=true, width=1\textwidth]{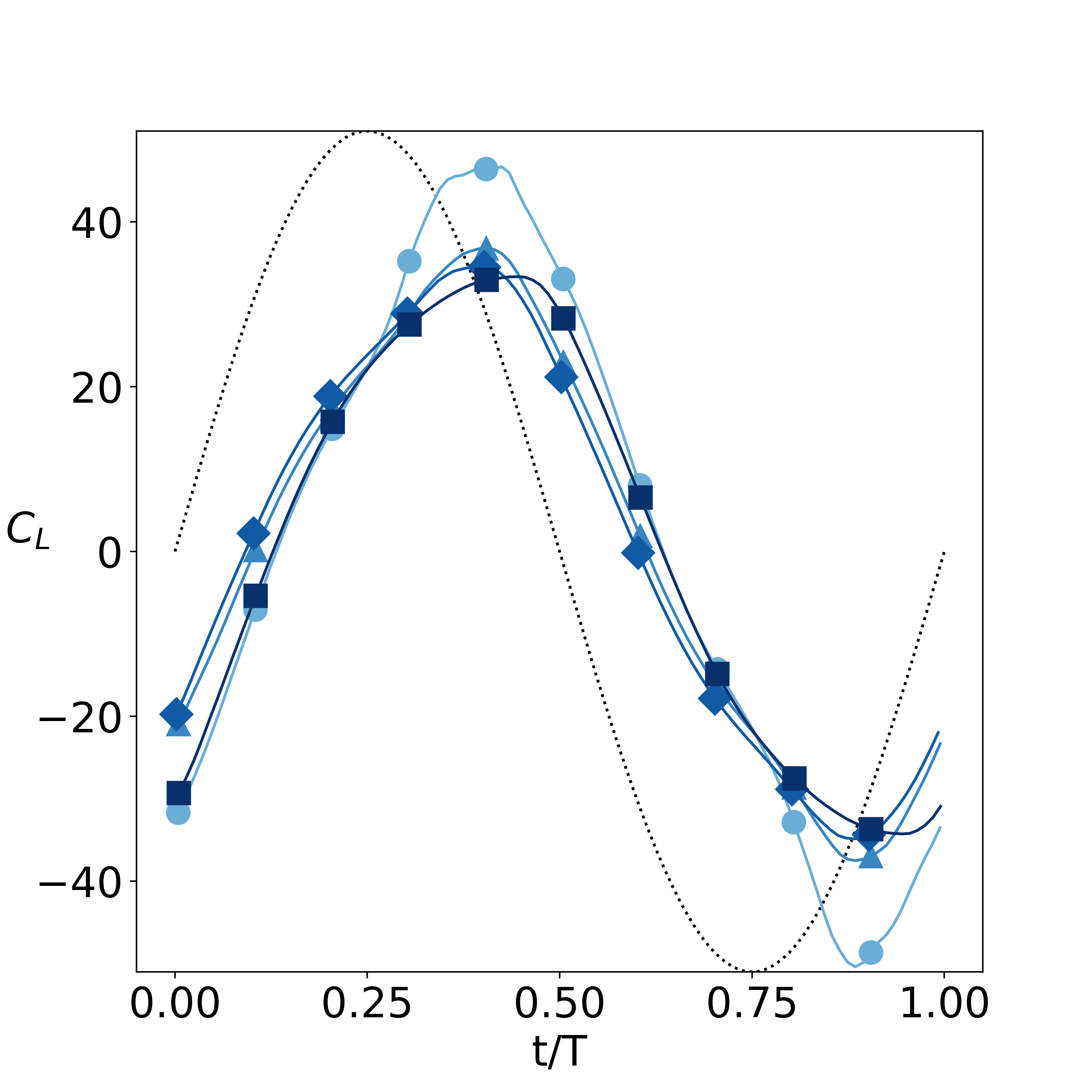}
		\subcaption{}
	\end{subfigure}
	\begin{subfigure}[hb]{.21\textwidth}
		\includegraphics[trim=1.cm 0 2.cm 1.8cm, clip=true, width=1\textwidth]{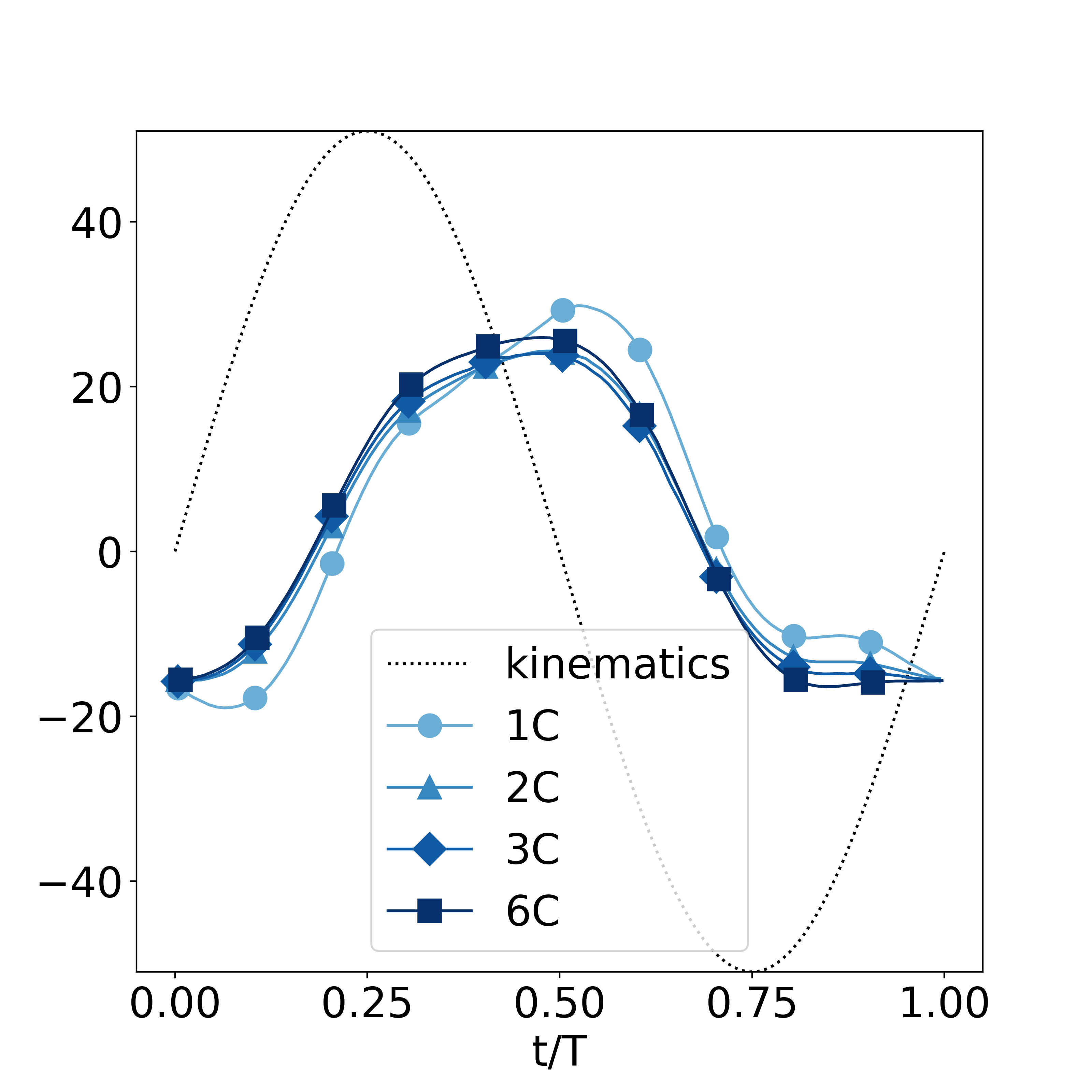}
		\subcaption{}
	\end{subfigure}
\caption{Phase-averaged lift coefficient $C_L$ evolution for various $S/C$ of foil undergoing (a) roll and (b) twist-roll motions over one cycle.}
\label{fig:tencycle}
\end{figure}

Finally, Fig.~\ref{fig:tencycle} shows the phase-averaged lift coefficient $C_L$ for all span lengths and both pure roll and twist-roll kinematics over a motion cycle. First, we note that the relative magnitude of the lift is substantially reduced when adding twist to the kinematics. This is because twist is out of phase with roll, reducing the angle of attack (Fig.~\ref{fig:aoa}) and therefore unloading the foil. Comparing the results within each kinematic case, we can see that despite the large differences in the flow and the $C_p$ distribution, the lift coefficients are quite similar. 

A clear difference in the force response shown in Fig.~\ref{fig:tencycle} is the large bumps in $C_L$ on the shortest foil at around $t/T=0.4,0.9$ during roll and $t=0.6,0.1$ during twist-roll. These large lift forces are due to the LEV remaining near the foil longer, and thereby reducing the local pressure substantially.

\subsection{Spanwise flow}

The previous section shows that foils with the strongest spanwise derivatives in their kinematics have stronger three dimensionality which increases the strength of their wake vorticities and increases the stability of their LEV, both of which increase pressure forces. Previous work in revolving wings suggests that spanwise flow helps stabilize and keep LEVs attached by draining their vorticity outboard \citep{Wong2015,Jardin2017} and we next investigate if spanwise flow is induced by the kinematic derivatives on 3D flapping wings. 

\begin{figure}
	\begin{subfigure}[hb]{.24\textwidth}
		\includegraphics[trim=0 0 2cm 2cm, clip=true, width=1\textwidth]{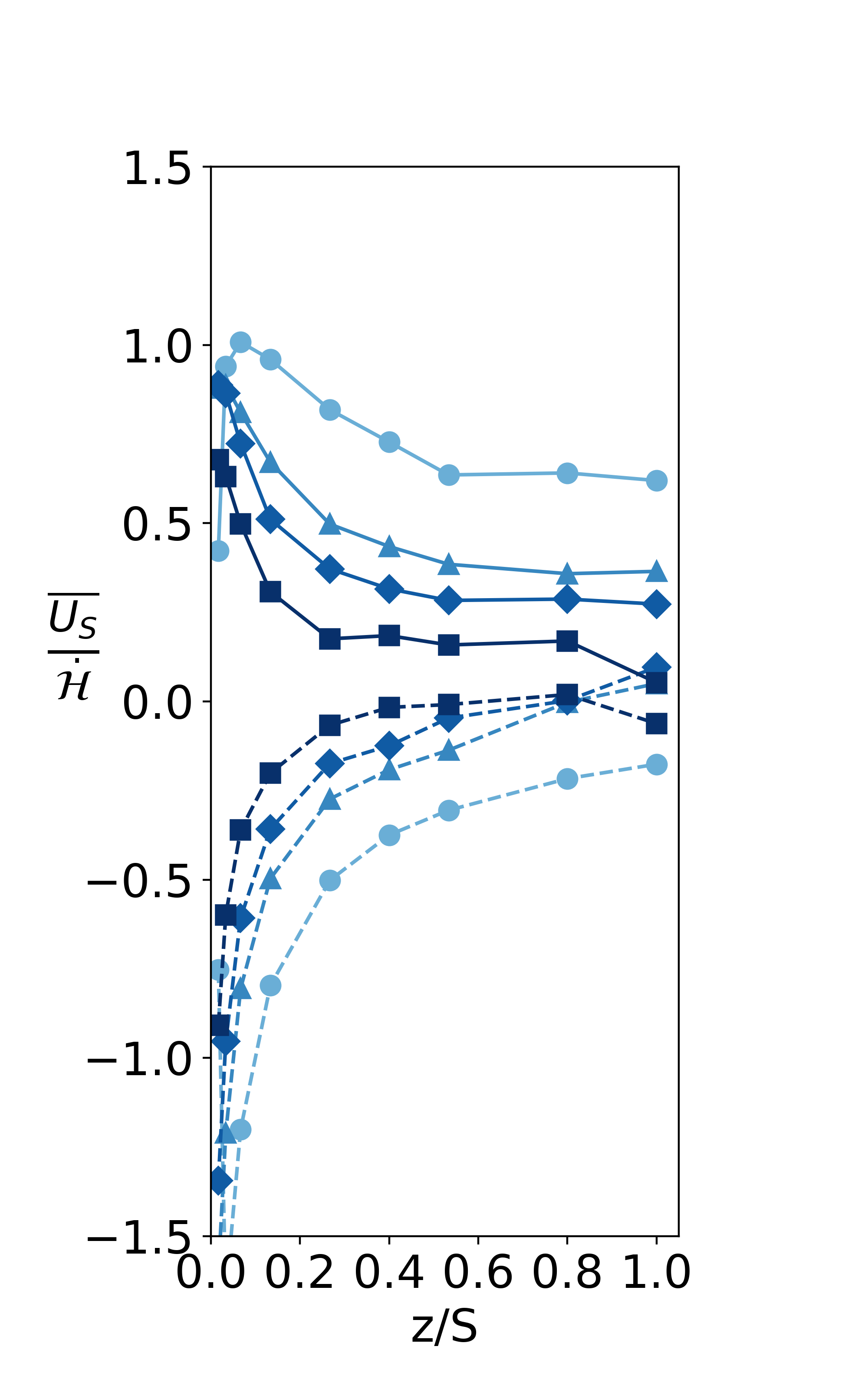}
		\subcaption{}
	\end{subfigure}
	\begin{subfigure}[hb]{.23\textwidth}
		\includegraphics[trim=0.5cm 0 2cm 2cm, clip=true, width=1\textwidth]{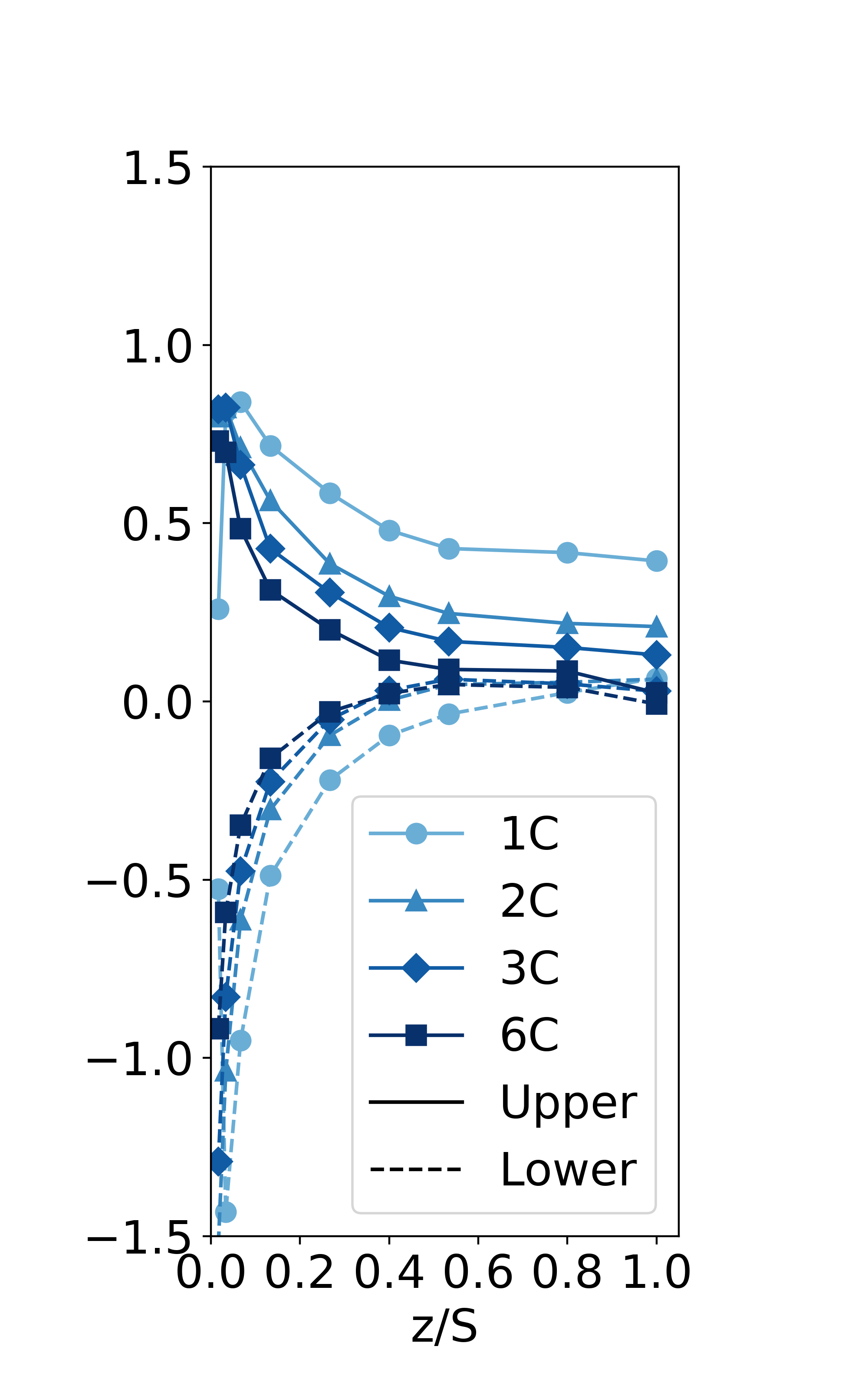}
		\subcaption{}
	\end{subfigure}
\caption{Sectional spanwise velocity normalized by the heave velocity at each $z/S$ section for (a) roll, and (b) twist-roll at $t/T=0.5$. Solid/dashed lines indicate flow on the upper/lower side of the foil. Positive values indicate flow towards the tip and vice versa.}
\label{fig:vspan}
\end{figure}

We define the sectional spanwise flow velocity over the foil surface as $\overline{U_{S}}^\pm=\frac{1}{C} \int_{0}^{C} U_{S}(x^\pm) dx$ where $U_S$ is the flow velocity tangent to the foil surface and perpendicular to the free stream direction, $x$ is the position on $C$, and $\pm$ represents either the velocity on the upper or lower surface of the section. Fig. \ref{fig:vspan} plots this quantity scaled by the maximum sectional heave velocity $\dot{\mathcal{H}}_{max}\propto z/S$ at each section along the span. 

The figure shows $\overline{U_S}$ is roughly proportional to $\dot{\mathcal{H}}_{max}\propto z/S$ from the mid span to the tip for both pure roll and twist-roll. As the span length increases (i.e. for higher \AR) this relatively flat region occurs earlier and the strength of the spanwise velocity decreases, both of which indicate a more 2D flow. Close to the root, both the heave velocity and spanwise velocity go to zero by symmetry, but their ratio is finite and $\sim 1$ . Also note the strength of the spanwise velocity for twist-roll is smaller than for pure roll due to relative decrease in the kinematic derivative in that case. The drop in spanwise velocity for $S=6C$ at $z=S$ shown in Fig.~\ref{fig:vspan} marks the location of strong vortex breakdown in Fig. \ref{fig:inst_roll} and indicates that the small spanwise derivative of the kinematics for this large aspect ratio case is not sufficient to overcome the spanwise vortex instabilities found on an infinite foil \citep{zurman2020}. 

\section{Discussion}

\begin{figure*}
\centering
	\begin{subfigure}[hb]{0.45\textwidth}
		\includegraphics[trim=0 0 1.5cm 0, clip=true, width=1\textwidth]{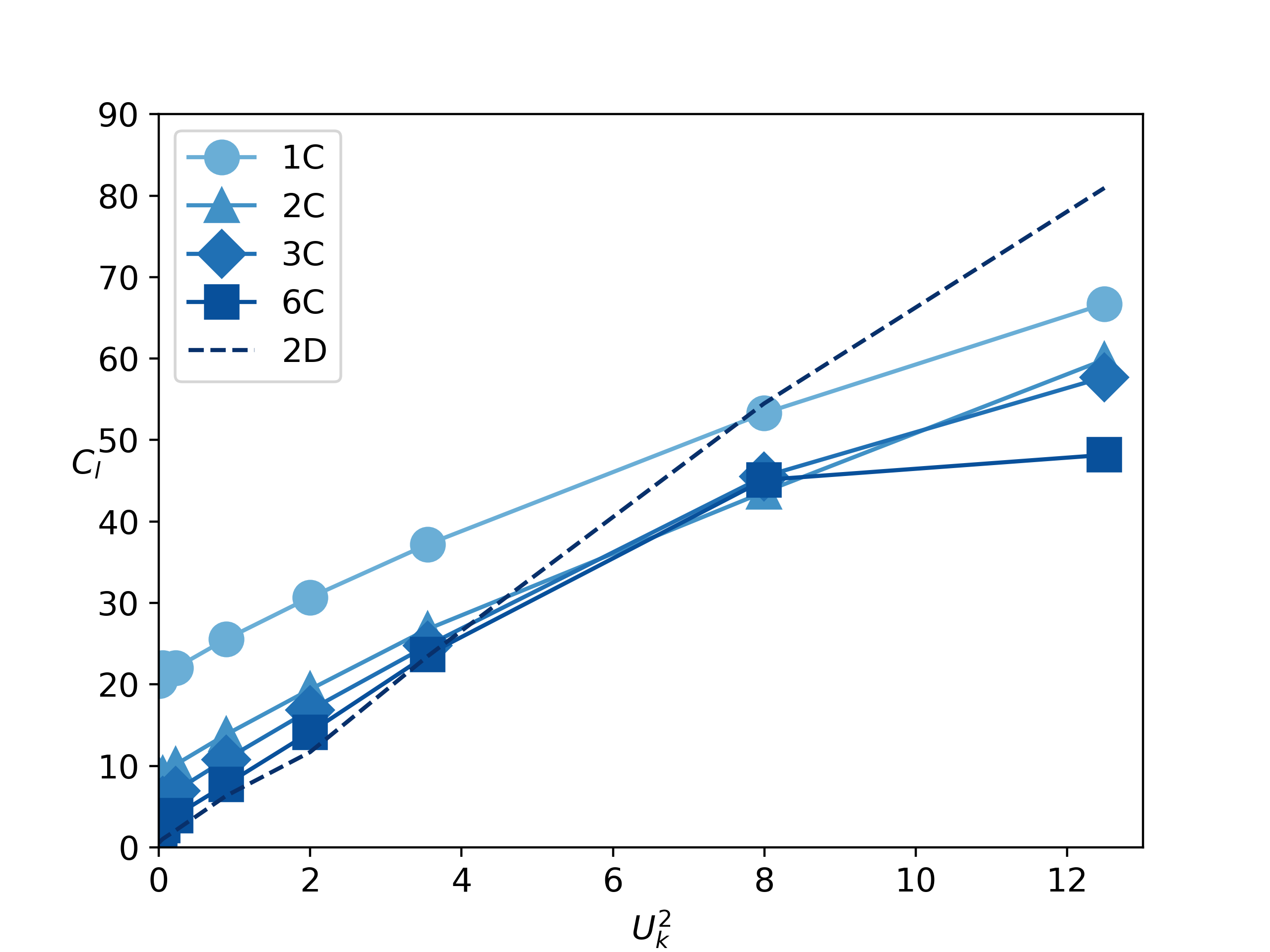}
		\subcaption{}
	\end{subfigure}
	\begin{subfigure}[hb]{0.45\textwidth}
		\includegraphics[trim=0 0 1.5cm 0, clip=true, width=1\textwidth]{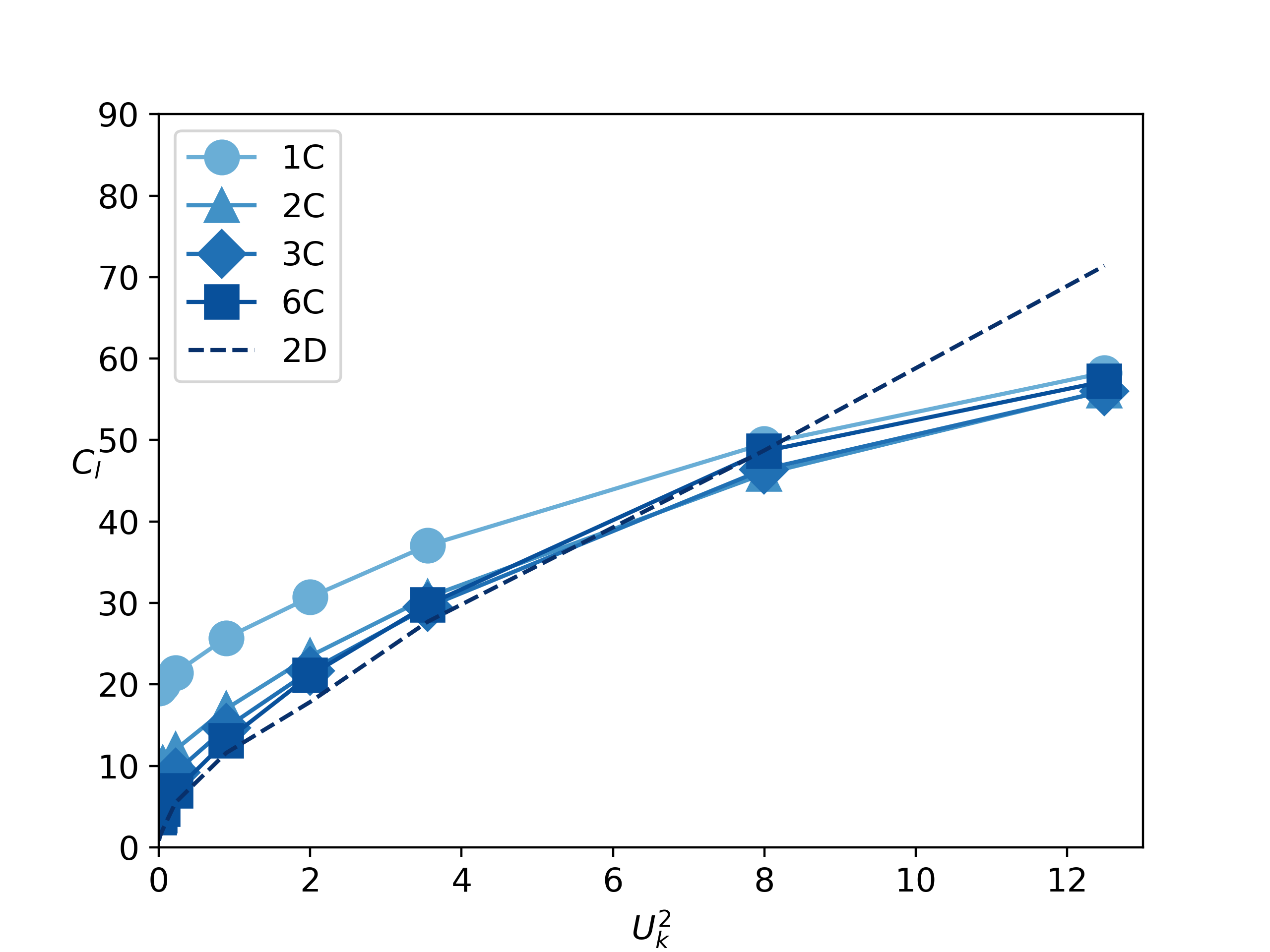}
		\subcaption{}
	\end{subfigure}
\caption{Peak sectional lift coefficients against $U_k^2=\dot{\mathcal{H}}_{max}^2/U_\infty^2$ for (a) roll and (b) twist-roll kinematics for 3D foils of different spans. Strip theory results from 2D foils with the same kinematics are also shown.}
\label{fig:cl}
\end{figure*}

\begin{table*}
 \caption{\label{tab:prandtl}3D slope ($a_{3D}$) change based on aspect-ratio compared to $a_{2D}$ for $t/T=0.5$. The slopes are measured using best-fit linear trend lines to the peak sectional lift from the root up to $z/S=0.8$. Data from $t/T=0$ produce the same trend (not presented here).}
 \centering
 \begin{tabular}{ccccccccc}
 \hline
 & &\multicolumn{3}{c}{Roll at $t/T=0.5$, $a_{2D}=6.52$}&&\multicolumn{3}{c}{Twist-roll at $t/T=0.5$, $a_{2D}=5.55$}\\
\cline{3-5}
\cline{7-9}
 $s/c$ & $AR_{eff}$ & $a_{3D}(pred)$ & $a_{3D}(sim)$ & $\left|\epsilon\right|$ (\%) && $a_{3D}(pred)$ & $a_{3D}(sim)$ & $\left|\epsilon\right|$ (\%) \\\hline
 6 & 7.2 & 5.07 & 4.06 & 24.8 && 4.46 & 4.47 &   0.2\\
 3 & 4.2 & 4.37 & 4.37 &   0.2 && 3.92 & 4.10 &   4.4\\
 2 & 3.2 & 3.97 & 4.13 &   3.9 && 3.59 & 3.81 &   5.9\\
 1 & 2.2 & 3.38 & 3.74 &   9.5 && 3.10 & 3.18 &   2.5\\
 \hline
 \end{tabular}
 \end{table*}

The spanwise derivative in the 3D flapping foil kinematics has been shown to induce a spanwise flow of proportional strength and this has a strong impact on the flow field. This spanwise flow overwhelms the documented behaviour of flapping infinite foils where the vortex strength is proportional to the scaled kinematic velocity $St_A$, and vortex instabilities leading to 3D structures are found at both very low and high $St_A$ \citep{zurman2020}. On a finite wing undergoing 3D kinematics, the spanwise flow helps distribute vorticity along the span, greatly increasing the strength of the vortices at the root and increasing the LEV stability for foils with short spans. Only the longest span length $S=6C$ shows similar behavior to infinite foils since the spanwise derivatives and spanwise flow are small. Nominally these values will falls to zero as the span length grows infinitely. Even for $S=6C$ the vortex structures are generally 3D and it is only safe to apply 2D strip theory to cross sections experiencing medium Strouhal number $St_A\sim0.3$.

However, the forces on the foil are much less sensitive to \AR\xspace than the wake, implying that it could possible for a simple model to modify 2D strip-theory sectional forces to approximately predict 3D finite foil loading. Consider the sectional lift coefficient $C_l(z/S) = \oint C_p n_y ds/C$, where $s$ is the 1D path around the $z/S$ foil section, such that $C_L\propto\int_0^S C_l dz/S$. Fig.~\ref{fig:cl} shows the peak $C_l$ at each section along the span of the finite wings as well as from 2D simulations with the same sectional kinematics. The kinematics at each value of $z/S$ match for all 3D and 2D simulations but the sectional coefficients do not, meaning the strip theory predictions must be modified to represent finite wing sectional forces.

The unsteady Bernoulli equation suggests that the peak sectional lift should scale with the peak kinetic energy of the foil section, i.e. $\frac 12 \rho \dot{\mathcal{H}}_{max}^2$. Fig.~\ref{fig:cl} tests this scaling by plotting peak $C_l$ against a normalized sectional kinetic energy $U_k^2=(\dot{\mathcal{H}}_{max}/U_\infty)^2\propto St_A^2$. We see that this scaling hold linearly for the 2D strip theory, and monotonically for the 3D sectional lift. Indeed, the slope of the peak sectional lift $a_{3D}=dC_L/d(U_k^2)$ changes with the aspect ratio, tending towards the strip theory slope as $S/C$ increases. This lift-curve slope, $a_{3D}$ can be determined using Prandtl's lifting line theory\cite{Anderson1995} as,
\begin{equation} \label{eq_prandtl}
	a_{3D}=\frac{a_{2D}}{1+a_{2D}/\pi \textrm{\AR}}
\end{equation}
where $a_{2D}$ is the lift-curve slope of the 2D case. This finite-length correction developed for steady flow works surprisingly well to correct the peak sectional lift on flapping foils, Tab.~\ref{tab:prandtl}. The only deviation greater than 10\% is the $S=6C$ roll case, where the lack of significant spanwise flow to stabilize the extremely strong tip vortices has greatly reduced the sectional lift on the end of the foil. In contrast, the corrected strip theory prediction have less than 6\% error for all the 3D foils undergoing twist-roll case because of its more moderate loading. 

\begin{figure}
    \frame{\includegraphics[width=0.45\textwidth]{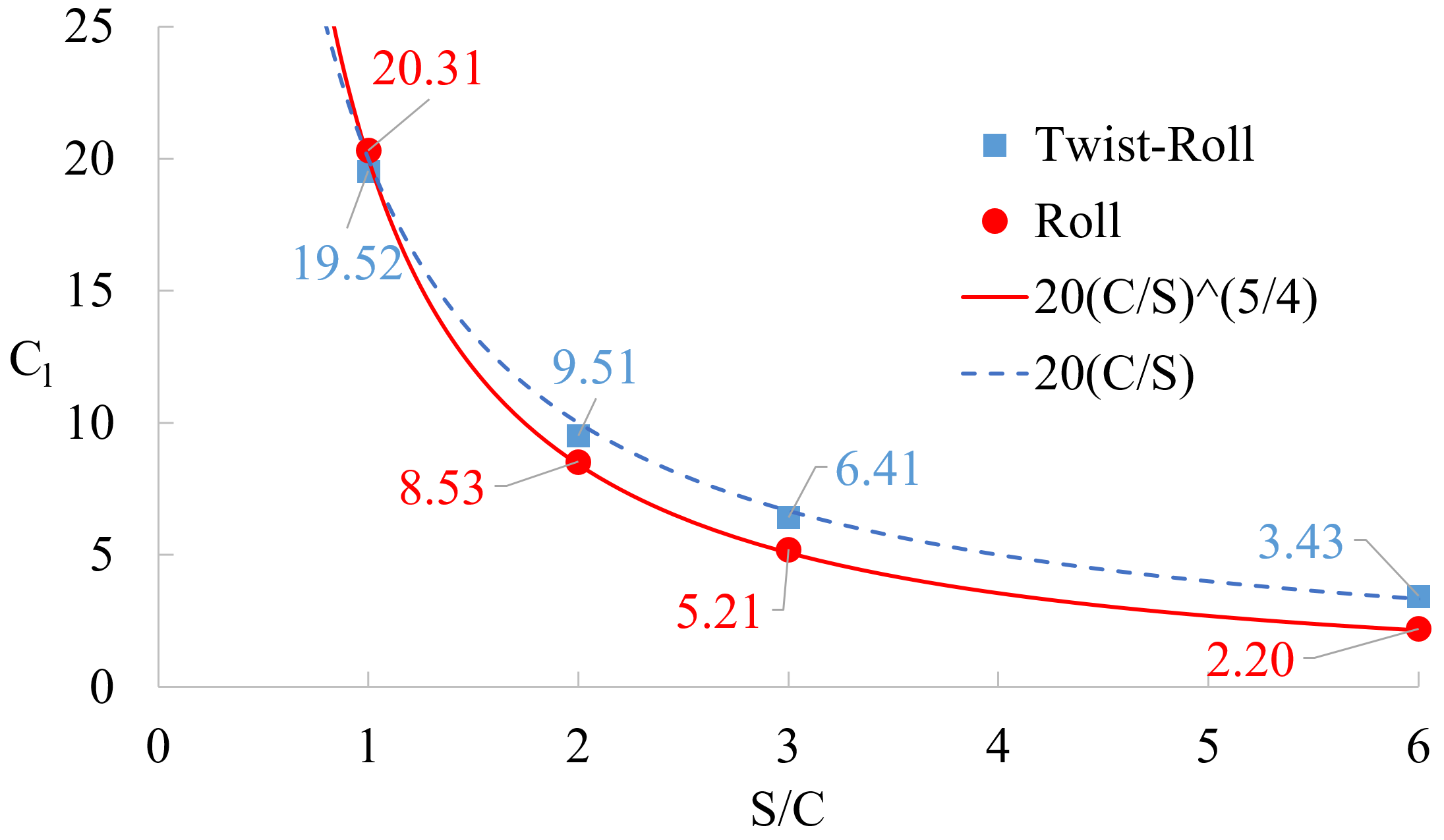}}
    \caption{Peak sectional $C_l$ at the root $z/S=0$.}
\label{fig:rootcl}
\end{figure}

In addition to the slope, the peak sectional $C_l$ at the root ($z=0$) also depends on $S/C$, as isolated in Fig.~\ref{fig:rootcl}. The forces for finite $S/C$ fit a simple power law extremely well, tending to zero as $S/C$ becomes large in agreement with the strip theory result.

\section{Conclusion}

In this work, we analyze the vortex structures and force characteristics of foils with prescribed rolling and twisting kinematics. Using three-dimensional Navier-Stokes simulations and matched kinematics per cross-section, we study the influence of increasing \AR s up to an idealized infinite span and compare it with the strip theory (2D analysis). If the \AR\xspace is sufficiently long, the flow induced by a finite foil undergoing 3D kinematics shows characteristics of the infinite foil flow. However, this flow is still generally 3D and strip theory is only applicable to the sections experiencing Strouhal numbers with 2D characteristics $St_A \approx 0.3$ \cite{zurman2020}. By decreasing the \AR, the spanwise derivative in the applied kinematics increases, inducing a strong spanwise flow on both upper and lower sides of the foil (in opposite directions) which promotes flow interaction between the sections. The shortest aspect ratio ($S/C=1$) therefore produces much stronger vortices, particularly at the root where they are 10-90 times stronger than the longest finite foil ($S/C=6$) depending on the kinematics applied.

Overall, we find that adding a $90^\circ$-lag twist can reduce the vortex breakdown produced by pure rolling motion near the tip. Twist also delays the LEV detachment. It indicates that small chordwise flexibility towards the tip, as seen in the animal's flipper or bird's wing, likely promotes more stable vortex structures and possibly improves force production.

The forces on the finite wings are also sensitive to \AR, but the impact is less extreme than on the flow structures. We find the peak sectional $C_l$ scales with local $St_A^2$, recognized as the added mass scaling \cite{Garrick1936,VanBuren2018}, and this scaling becomes linear in the limit of infinite \AR. The slope $ dC_l/d(St_A^2)$ surprisingly follows the Prandtl finite-wing trend, i.e. lower slope for low \AR\xspace and vice versa. This aspect ratio scaling has been reported to fit the circulatory force by varying aspect ratios \cite{Ayancik2019,Moored2018,Green2008}, but this scaling has not been previously used to scale 3D forces from 2D strips. We find strip theory predictions with an \AR\xspace correction have error below 6\% for a finite wing undergoing twist-roll motion, indicating it is possible to scale the sectional forces experienced by 3D foils based on results from 2D strips using a version of Prandtl finite-wing theory. 

\section*{Acknowledgments}
We would like to thank Indonesia Endowment Fund for Education (LPDP), IRIDIS High Performance Computing Facility with its associated support services at the University of Southampton and the Office of Naval Research Global award N62909-18-1-2091 for the completion of this work.

\appendix*
\section*{Appendix: grid convergence}
\begin{figure*}[hbt!]
\centering
	\begin{subfigure}[b]{.45\textwidth}
		\includegraphics[trim=0 0 0 1.2cm, clip=true, width=1\textwidth]{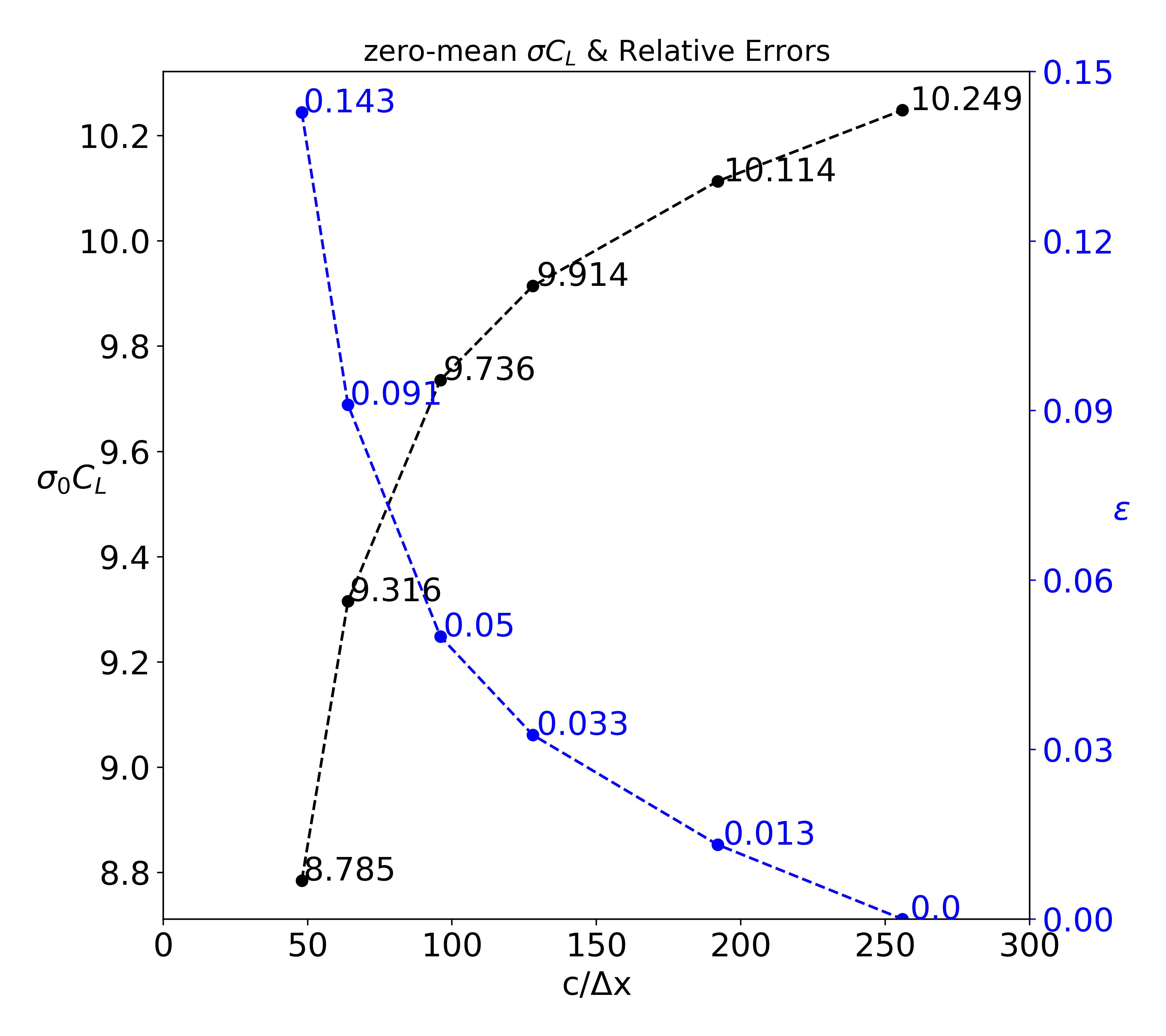}
		\subcaption{}
	\end{subfigure}
    \begin{subfigure}[b]{.45\textwidth}
		\includegraphics[trim=0 0 0 1.2cm, clip=true, width=1\textwidth]{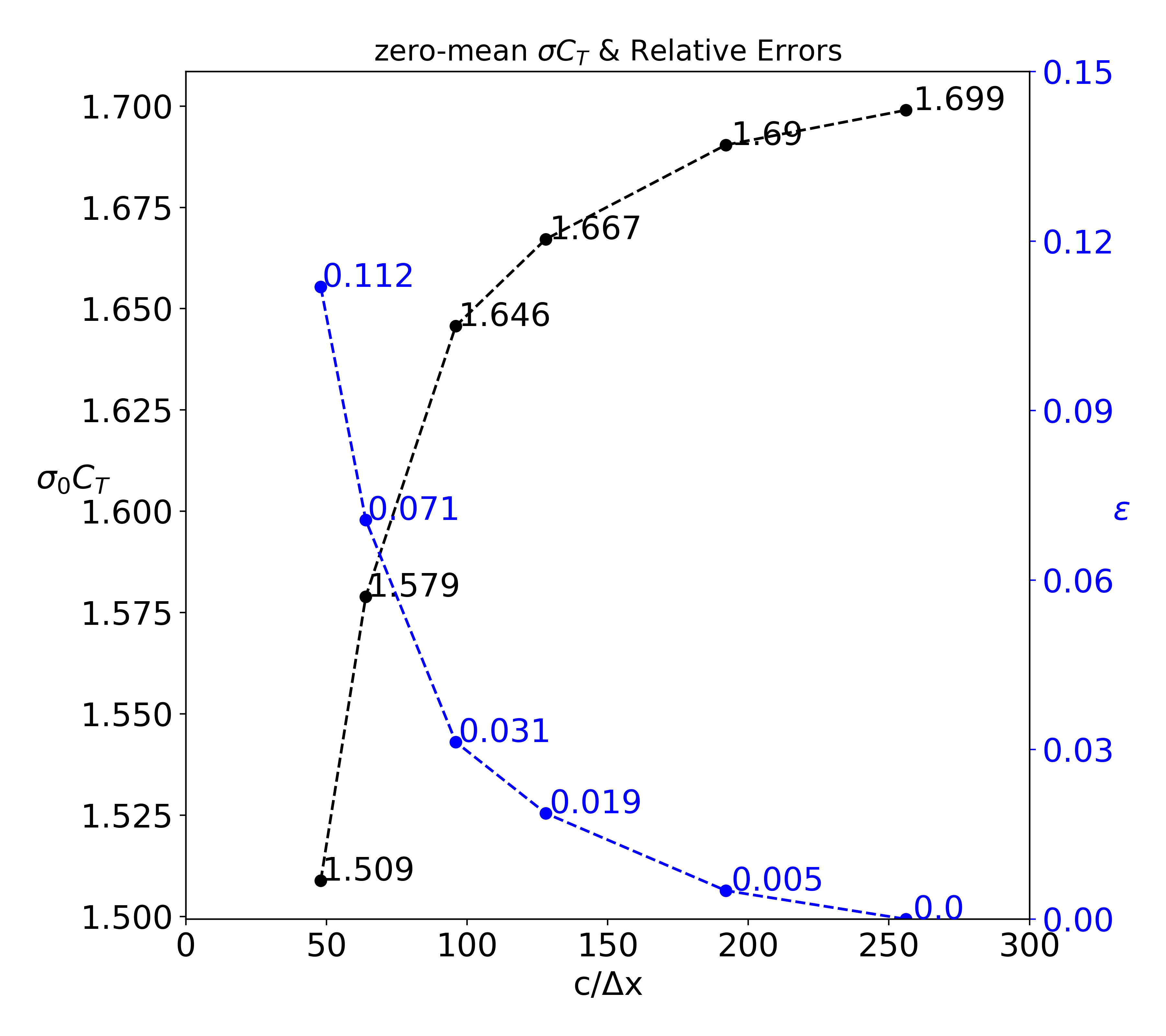} 
		\subcaption{}
	\end{subfigure}
	\begin{subfigure}[b]{.45\textwidth}
		\includegraphics[trim=0 0 0 1.2cm, clip=true, width=1\textwidth]{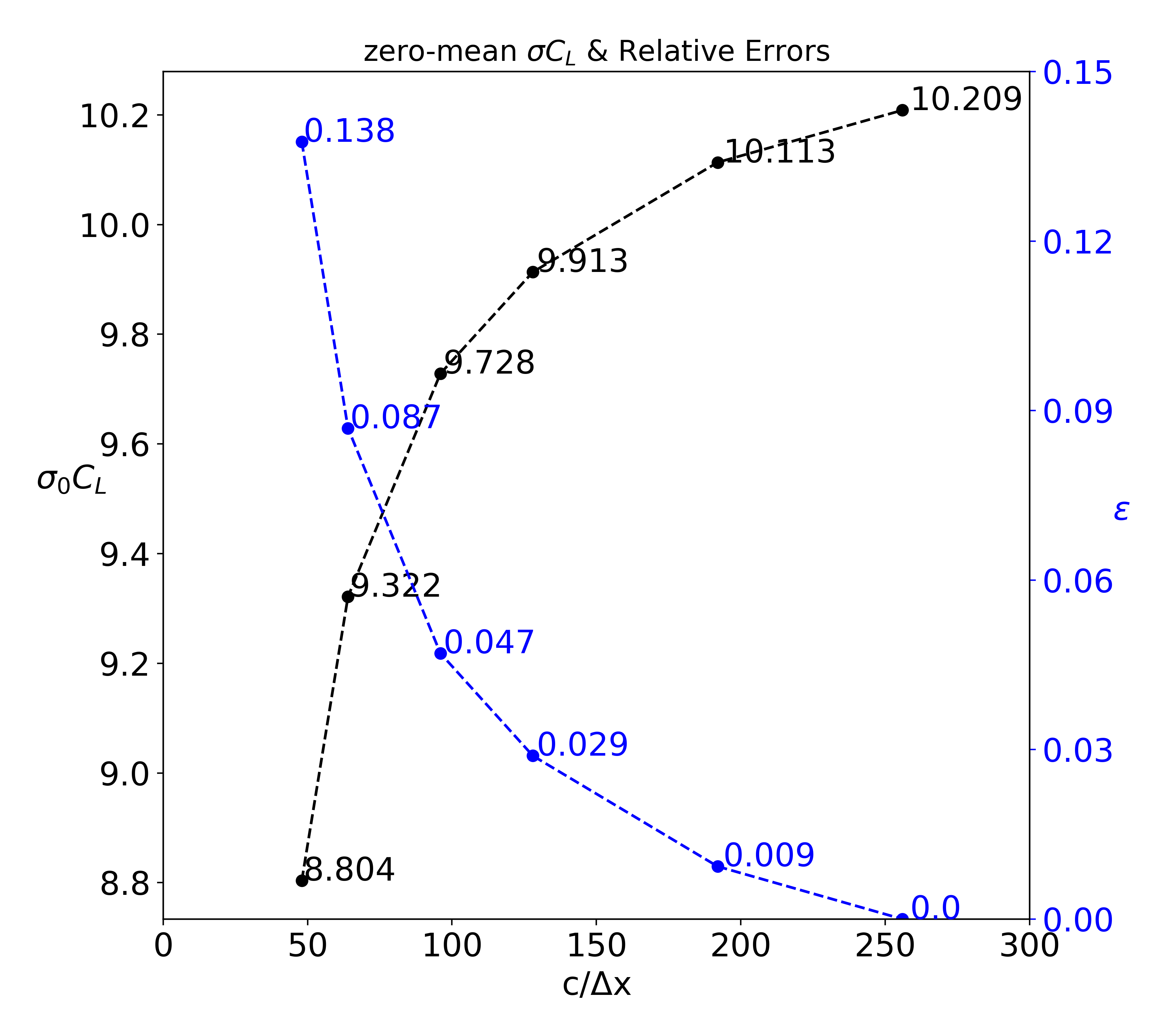}
		\subcaption{}
	\end{subfigure}
    \begin{subfigure}[b]{.45\textwidth}
		\includegraphics[trim=0 0 0 1.2cm, clip=true, width=1\textwidth]{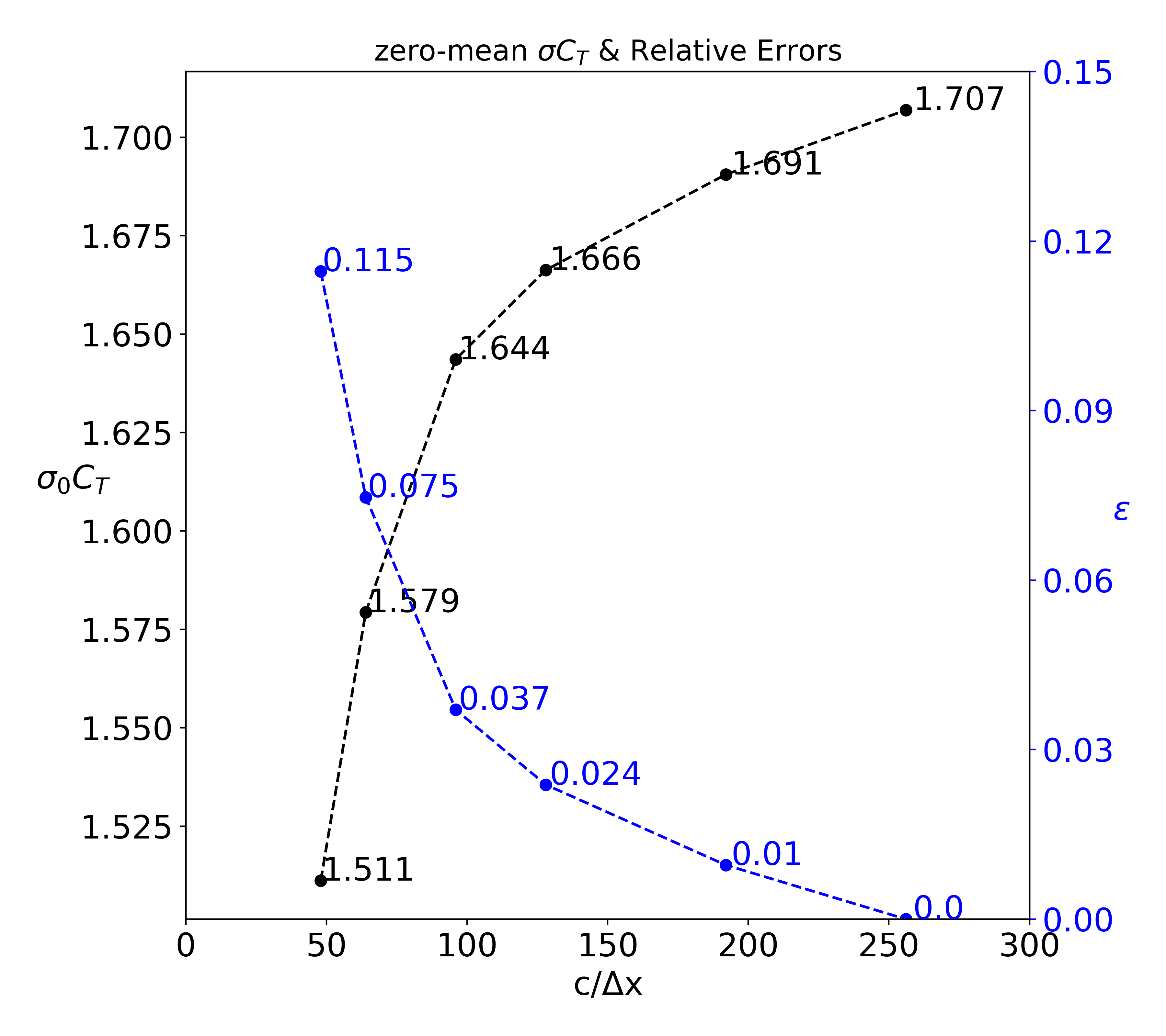} 
		\subcaption{}
	\end{subfigure}
\caption{Grid convergence statistics for 3D and 2D simulations of infinite foils in a combination of heaving-pitching motion at \(Re =5300\), \(A_D=1.0\), $St_D=0.3$ and $\theta_{bias}=10^\circ$. The statistics are presented in zero-mean standard deviation $\sigma_0$ for (a) 2D $C_L$, (b) 2D $C_T$, (c) 3D $C_L$, and (d) 3D $C_T$. $\epsilon$ is relative error to $C/\Delta x=256$.}
\label{fig:conv}
\end{figure*}

\bibliography{Reference}

\end{document}